\documentclass[3p,referee]{elsarticle}
\usepackage{amssymb}

\usepackage{graphics}
\usepackage{color}
\usepackage{units}
\usepackage{subfigure}
\usepackage{epstopdf}
\usepackage{amsmath}

\usepackage{longtable}

\definecolor{darkred}{rgb}{0.5,0,0}
\definecolor{darkgreen}{rgb}{0,0.5,0}
\definecolor{darkblue}{rgb}{0,0,0.5}
\definecolor{violett}{rgb}{0.5,0,0.5}
\definecolor{pink}{rgb}{1,0,1}

\makeatletter
  \renewcommand{\fps@figure}{htb}
  \renewcommand{\fps@table}{htb}
\makeatother

\def\apj{\emph{ApJ.}}
\def\aj{\emph{AJ.}}
\def\aap{\emph{A.\& A.}}
\def\apjs{\emph{ApJS}}

\def\apjl{\emph{ApJ. Lett.}}

\begin{document}

\begin{frontmatter}

\title{Milagro Observations of  Potential TeV Emitters}

\author[label1,label16]{A.~A.~Abdo}
\author[label1]{A.~U.~Abeysekara}
\author[label2,label17]{B.~T.~Allen}
\author[label3,label18]{T.~Aune}
\author[label1,label19]{A.~S.~Barber}
\author[label4]{D.~Berley}
\author[label4]{J.~Braun} 
\author[label2]{C.~Chen}
\author[label5]{G.~E.~Christopher}
\author[label6]{T.~DeYoung}
\author[label7]{B.~L.~Dingus}
\author[label8]{R.~W.~Ellsworth}
\author[label9]{M.~M.~Gonzalez}
\author[label4]{J.~A.~Goodman}
\author[label10]{E.~Hays}
\author[label7]{C.~M.~Hoffman}
\author[label11]{P.~H.~H\"untemeyer}
\author[label7]{A.~Imran}
\author[label5]{B.~E.~Kolterman}
\author[label1]{J.~T.~Linnemann}
\author[label10]{J.~E.~McEnery}
\author[label12]{T.~Morgan}
\author[label5]{A.~I.~Mincer}
\author[label5]{P.~Nemethy}
\author[label7]{J.~Pretz}
\author[label12]{J.~M.~Ryan}
\author[label3,label15]{P.~M.~Saz~Parkinson}
\author[label13]{M.~Schneider}
\author[label14]{A.~Shoup}
\author[label7]{G.~Sinnis}
\author[label4]{A.~J.~Smith}
\author[label4,label21]{V.~Vasileiou}
\author[label7,label20]{G.~P.~Walker}
\author[label3]{D.~A.~Williams}
\author[label2]{G.~B.~Yodh}

\cortext[corresponding]{Corresponding author: udaraabeysekara@yahoo.com}

\address[label1]{Department of Physics and Astronomy, Michigan State University, BioMedical Physical Sciences Building, East Lansing, MI 48824}
\address[label2]{Department of Physics and Astronomy, University of California, Irvine, CA 92697}
\address[label3]{Santa Cruz Institute for Particle Physics, University of California, 1156 High Street, Santa Cruz, CA 95064}
\address[label4]{Department of Physics, University of Maryland, College Park, MD 20742}
\address[label5]{Department of Physics, New York University, 4 Washington Place, New York, NY 10003}
\address[label6]{Department of Physics, Pennsylvania State University, University Park, PA 16802}
\address[label7]{Group P-23, Los Alamos National Laboratory, P.O. Box 1663, Los Alamos, NM 87545}
\address[label8]{Department of Physics and Astronomy, George Mason University, 4400 University Drive, Fairfax, VA 22030}
\address[label9]{Instituto de Astronom\'ia, Universidad Nacional Aut\'onoma de M\'exico,D.F., M\'exico, 04510}
\address[label10]{NASA Goddard Space Flight Center, Greenbelt, MD 20771}
\address[label11]{Department of Physics, Michigan Technological University, Houghton, MI 49931, USA}
\address[label12]{Department of Physics, University of New Hampshire, Morse Hall, Durham, NH 03824}
\address[label13]{University of California Santa Cruz, Natural Science 2, 1156 High Street, Santa Cruz, CA 95064, USA}
\address[label14]{Ohio State University, Lima, OH 45804}
\address[label15]{Department of Physics, The University of Hong Kong, Pokfulam Road, Hong Kong, China}
\address[label16]{Current address: Operational Evaluation Division, Institute for Defense Analyses, 4850 Mark Center Drive, Alexandria, VA 22311-1882}
\address[label17]{Current address: Harvard-Smithsonian Center for Astrophysics, Cambridge, MA 02138}
\address[label18]{Current address: Department of Physics and Astronomy, University of California, Los Angeles, CA 90095}
\address[label19]{Current address: Department of Physics, University of Utah, Salt Lake City, UT 84112}
\address[label20]{Current address: National Security Technologies, Las Vegas, NV 89102}
\address[label21]{Current address: Laboratoire Univers et Particules de Montpellier, Universit\'e Montpellier 2, CNRS/IN2P3,  CC 72, Place Eug\`ene Bataillon, F-34095 Montpellier Cedex 5, France}

\begin{abstract}
This paper reports the results from three targeted searches of Milagro TeV sky maps: two extragalactic point source lists and one pulsar source list. 
The first extragalactic candidate list consists of 709 candidates selected from the \textit{Fermi-LAT} 2FGL catalog. The second extragalactic candidate list contains 31 candidates selected from the TeVCat source catalog that have been detected by imaging atmospheric Cherenkov telescopes (IACTs). In both extragalactic candidate lists Mkn 421 was the only source detected by Milagro. 
This paper presents the Milagro TeV flux for Mkn 421 and flux limits for the brighter \textit{Fermi-LAT} extragalactic sources and for all TeVCat candidates. The pulsar list extends a  previously published Milagro targeted search for Galactic sources. 
With the 32 new gamma-ray pulsars identified in 2FGL, the number of pulsars that are studied by both \textit{Fermi-LAT} and Milagro is increased to 52. 
In this sample, we find that the probability of Milagro detecting a TeV emission coincident with a pulsar increases with the GeV flux observed by the \textit{Fermi-LAT} in the energy range from 0.1 GeV to 100 GeV.
\end{abstract}

\begin{keyword}
astroparticle physics \sep pulsars \sep galaxies \sep active galactic nuclei \sep gamma-rays
\end{keyword}

\end{frontmatter}
\section{Introduction}
The Milagro gamma-ray observatory was a water Cherenkov detector located near Los Alamos, New Mexico, USA at latitude
$35.9^{\circ}$ north, longitude $106.7^{\circ}$ west and altitude 2630 m \citep{NewCrabPaper}.
Milagro recorded data from 2001-2008 and was sensitive to extensive air showers initiated by gamma-rays with energies from a few hundred GeV to $\sim$100 TeV. Unlike atmospheric Cherenkov
telescopes, Milagro had a wide field of view and it was able to monitor the sky with a high duty cycle($>90\%$).

The Milagro collaboration has performed blind source searches and found
a number of TeV sources (\cite{2004ApJ...608..680A} and \cite{Abdo2007} We refer to this as Milagro Galactic Plane Surveys). 
Blind searches for excess events over the full sky have a high probability of picking up random fluctuations.
Therefore, after trials correction, a full sky blind search is less sensitive than searches using a smaller predefined list of candidates.
The \textit{Fermi Large Area Telescope (Fermi-LAT)} collaboration published such a list known as the Bright Source List or 0FGL list \citep{2009yCat..21830046A}. 
In a previous publication, Milagro reported a search using 0FGL sources identified as Galactic sources \citep{2009ApJ...700L.127A}, which we will refer to as the Milagro 0FGL paper.

We report here two Milagro targeted searches for extragalactic sources. 
The first extragalactic candidate list is compiled from the extragalactic sources in the 2FGL catalog\citep{FermiLAT2FGLPaper}. 
The analysis presented in this paper looks for the TeV counterparts of these sources. 
The second extragalactic candidate list is made from the TeVCat catalog \citep{2008ICRC....3.1341W} of extragalactic sources.  
While TeVCat detections may include transient states of variable extragalactic sources, this search looks for long-term time averages by integrating over the full Milagro data set.
However, it is not appropriate to use the second extragalactic candidate list to perform a population study as it has candidates detected from several instruments with different sensitivities.

Our previous Milagro 0FGL publication  found that the \textit{Fermi-LAT} bright sources that were measured at or above 3 standard deviations in significance (3$\sigma$) by Milagro were dominated by pulsars and/or their associated pulsar wind nebulae (PWN). Therefore, in this paper we extend the previous Galactic search by making a candidate list from the pulsars in the 2FGL source list, and search for TeV emission from the sky locations of gamma-ray pulsars detected by the \textit{Fermi-LAT}. 
The angular resolution of Milagro $\left( 0.35^\circ< \delta\theta<1.2^\circ \right)$  is not sufficient to distinguish the PWN from the pulsar.

\section[]{Methodology} \label{methodology}

\subsection{Construction Of Candidate Lists}\label{list}

The first candidate list, which will be referred to as the 2FGL Extragalactic List, is derived from
the 2FGL catalog by looking for sources off the Galactic plane
($|b| > 10^\circ$) that have no association with pulsars.
There are 709 \textit{Fermi-LAT} sources within Milagro's sky coverage ($-7^\circ<$DEC$<80^\circ$), of which 72\% are associated with blazars. 
Among these blazars 4 are firmly identified as BL Lac\footnote{BL Lac is a type of active galaxy of known to be strongly $\gamma$-ray emitting objects \citep{2002A&A...384...56C}. } blazars and 12 are firmly identified
as FSRQ\footnote{Flat Spectrum Radio Quasar} type of blazars.

The second extragalactic candidate list, which we will  call the TeVCat Extragalactic List, is taken from TeVCat, an online gamma-ray source catalog (http://tevcat.uchicago.edu). As of February 8th, 2012 it contained 135 sources, of which 31 were located off the Galactic plane and
within Milagro's sky coverage.  
These 31 sources were all detected with Cherenkov telescopes and 23 are identified as BL Lac objects.

There are 52 sources in the 2FGL catalog associated with pulsars which are in the Milagro's sky coverage.  
Twenty of these pulsars were already considered as candidates in the Milagro 0FGL publication. 
So the third candidate list, which will be called the Pulsar List, consists of only the 32 new pulsars. 
Of these, 17 were identified as pulsars by
pulsations seen in \textit{Fermi-LAT} data and the
remaining 15 sources were labeled as pulsars in 2FGL because of their spatial
association with known pulsars.

\subsection{Spectral Optimizations}\label{SpectralAssumptions}
In order to optimize the sensitivity to photon sources, Milagro sky maps are constructed by plotting the location for each event with a weight based on the relative probability of it being due to a primary photon or hadron \citep{NewCrabPaper}. The weight calculation depends on the assumed photon spectrum and can be suboptimal (but not incorrect) if the weight optimization hypothesis is considerably different from the actual source spectrum. 
The weights are therefore optimized separately for two hypotheses.

For the extragalactic candidate lists, a power law with spectral index $\alpha$= -2.0 with a 5 TeV exponential cut-off ($E^{-2.0} e^{- \frac{E}{5 \rm{TeV}}}$) was assumed. 
This choice reflects the fact that  when TeV gamma-rays travel cosmological distances they are attenuated due to interactions with photons from the extragalactic background light
\citep{2005ApJ...618..657D} with the result that the energy spectrum of extragalactic sources cut off at high energies.  
This spectral assumption is also similar to the power law spectral index and the cut-off energy measured for Mkn 421 and Mkn 501 by \cite{2001ApJ...560L..45K}. 
However, the choice of 5 TeV cut off might reduce the sensitivity of Milagro to the AGNs with lower cut off energies.
For the Pulsar List, a power law with spectral index $\alpha = -2.6$ with no TeV cut-off is used, as was done for the previous Milagro 0FGL and Galactic Plane Survey papers.

\subsection{Source Detection Technique}

The expected significance at a sky location with no true emission is a Gaussian random variable with mean 0 and unit standard deviation \cite{2004ApJ...608..680A}.  
A common treatment of N candidate searches is to use a trials correction technique.
Here one choose a significance threshold, calculate the tail probability (p-value) $\lambda$, and adjusts the p-value threshold to $\frac{\lambda}{N}$. The purpose of the trials correction is to maintain, at the value $\lambda$, the probability of a background fluctuation producing one or more false discoveries among the N searches.\\

The False Discovery Rate (FDR) technique discussed in \cite{Christopher2001} offers some advantages over the trials correction technique.  Instead of controlling the expected probability of having even one false detection, FDR controls the \textit{expected fraction} of false discoveries among a set of detections; that is, it controls the contamination fraction of the \textit{lists} of associations, rather than the probability of a random individual association being accepted\footnote{The required calculations are quite simple and can be implemented in a spreadsheet after the significances of the searches on a list are calculated.}. The key input parameter is again a probability $\lambda$, but now $\lambda$ represents the expected fractional contamination of any announced set of detections.  Based on this input parameter, the method dynamically adjusts the detection threshold but in a way that depends on the properties of the entire list of search significances (converted into p-values).  This dynamic adjustment is sensitive to whether the distribution of p-values is flat (as would be expected if there were no detectable sources) or skewed to small p-values (i.e. large significances).  This adjustment lowers the significance threshold for detection if a list is a ``target-rich environment'' in such a way that the expected fraction of false discoveries among the announced detections remains at the fraction
$\lambda$.  In particular, the most significant candidate is required to have a p-value of
$\lambda/N$ just as in the trials-correction method, but the $n$-th most significant candidate need only have a p-value less than $\lambda \times n/N$.  As a result, this technique has a higher efficiency for finding real detections, while producing the same results as a trials-correction method in target-poor environments where the only decision is whether to report zero or one detections.  The method adjusts for both the length of the search list and the distribution of the significances found within the search lists.  However, we note that as a result, a given candidate location might pass the FDR criteria on one search list, but fail in another. We also emphasize that $\lambda$ controls the \textit{expected} contamination, \textit{i.e.} averaged over potential lists of associations, not the contamination fraction on a specific list\footnote{For example, in an environment with no real sources, one expects to report an empty list $(1-\lambda\ )\times100 \%$ of the time, and about $\lambda\times100 \%$ of the time one would report a list  having at least a single (false) candidate}. The reader is referred to \cite{Christopher2001} for further details of the method\footnote{We assume that the search points are uncorrelated, as the angular separations between target locations are normally much more widely seperated than the Milagro point spread function.}.
In the \cite{2009ApJ...700L.127A} paper for the Galactic-oriented search with N = 35, a criterion of $3 \sigma$ was used but it was also found that an FDR criterion of $\lambda=0.01$ produced the same list of associations. Specifying $\lambda$ rather than a $\sigma$ threshold also allows using a single criterion for treating each of the search lists. The analyses presented in this paper uses $\lambda=0.01$ for defining a TeV association for all our search lists, but significance thresholds are also tabulated in Table~\ref{tab:Summary} for $\lambda= 0.1,0.05,$ and $0.001$ so that readers can choose the potential contamination level of candidate lists. Specific candidates passing looser cuts are also denoted as footnotes to the search list tables.

\subsection{Stacking Methodology}\label{sec:StackingMethodology}
The FDR technique can be used to search for individual candidates with a TeV association. A stacking analysis can be used to search for evidence of collective TeV emission among the undetected candidates by studying their mean flux. This paper uses the stacking methodology of Section 3 in \cite{Stacking}. The significance of the stacked flux is given by Equation~\ref{Signif} below.

\begin{equation}\label{Signif}
\rm{Significance} =  \frac{\left< I \right>}{\sqrt{\vee{\left(  \left< I \right> \right)}}},
\end{equation}
where $\left< I \right>$ is the weighted average flux as defined in Equation~\ref{AvgFlux} below and $ \vee{ \left< I \right>}$ is its variance, defined in Equation~\ref{VarFlux}.

\begin{equation}\label{AvgFlux}
\left< I \right> = \frac{\sum{\frac{I_i}{\sigma_i^2}}}{\sum{\frac{1}{\sigma_i^2}}}
\end{equation}

\begin{equation}\label{VarFlux}
 \vee{ \left< I \right>} = \frac{\sum{\frac{1}{\sigma_i^2}}}{\sum{ \left( \frac{1}{\sigma_i^2} \right)^2}}
\end{equation}
Here $I_i$ is the flux of each candidate and $\sigma_i$ is the standard deviation of flux of each candidate.

\subsection{Flux Calculation}
The flux calculation involves a convolution of the Milagro effective area as a function of energy using an assumed energy spectrum, so the flux has some dependence on the assumed energy spectrum. This dependence is greatly reduced when the flux is calculated at the median energy of the detected gamma-ray events at the declination of a source \citep{2009ApJ...700L.127A}.  Therefore, we report the flux at approximately the median energy. Using a similar argument to that in the Milagro 0FGL paper, the flux is derived at 35 TeV for the Pulsar List. For the extragalactic spectral assumption the median energy varies between 6 and 11 TeV, and we choose 7 TeV to report the flux for extragalactic source candidates.

\begin{figure}
\centering
\includegraphics[width=80mm]{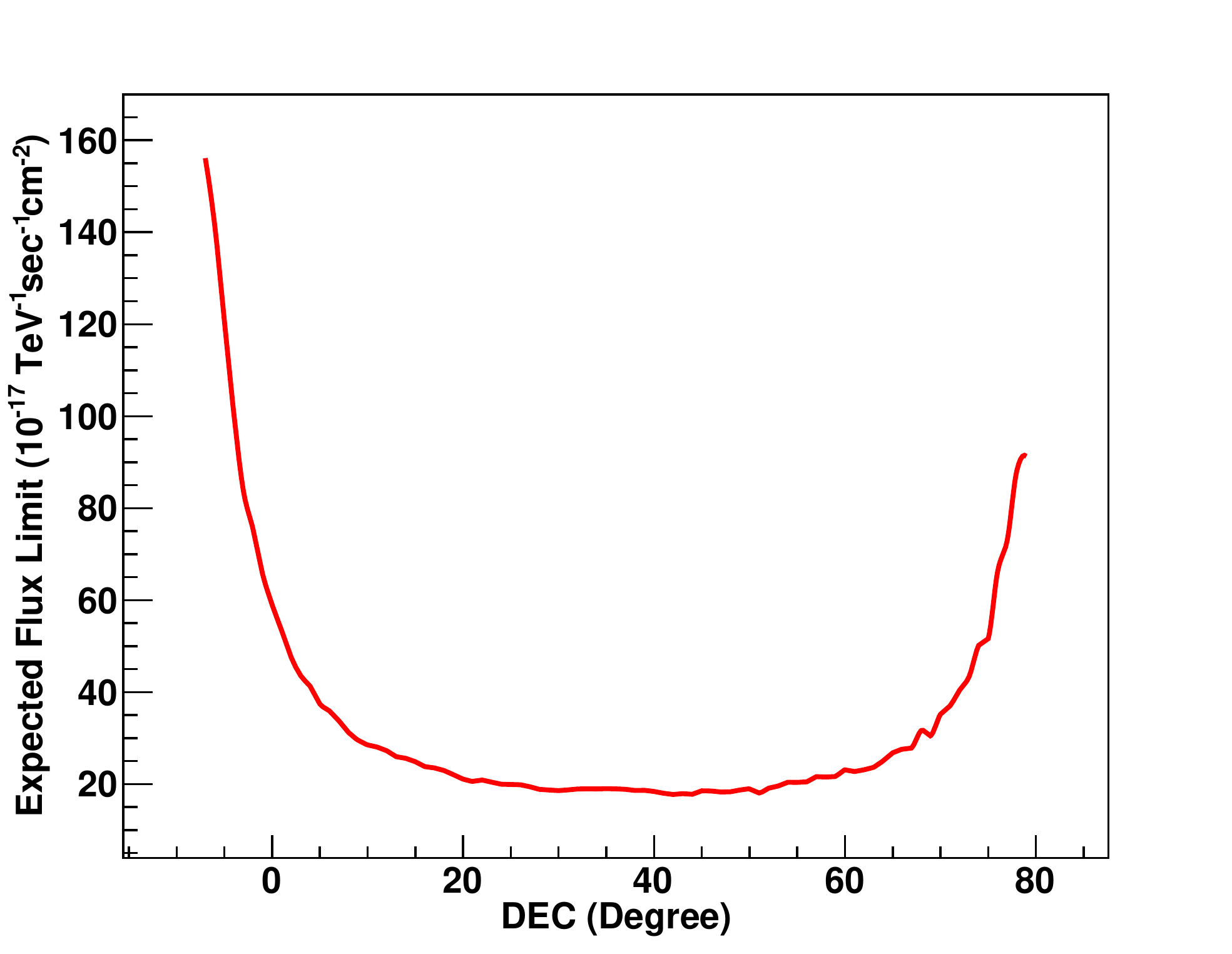}
\caption{The expected 95\% confidence level flux upper limit for Galactic sources corresponding to zero excess derived at 35 TeV for each declination band of the Milagro sky maps made with spectral assumption $dN/dE \propto E^{-2.6}$.}\label{UpperLimit2p6}
\end{figure}

\begin{figure}
\centering
\includegraphics[width=80mm]{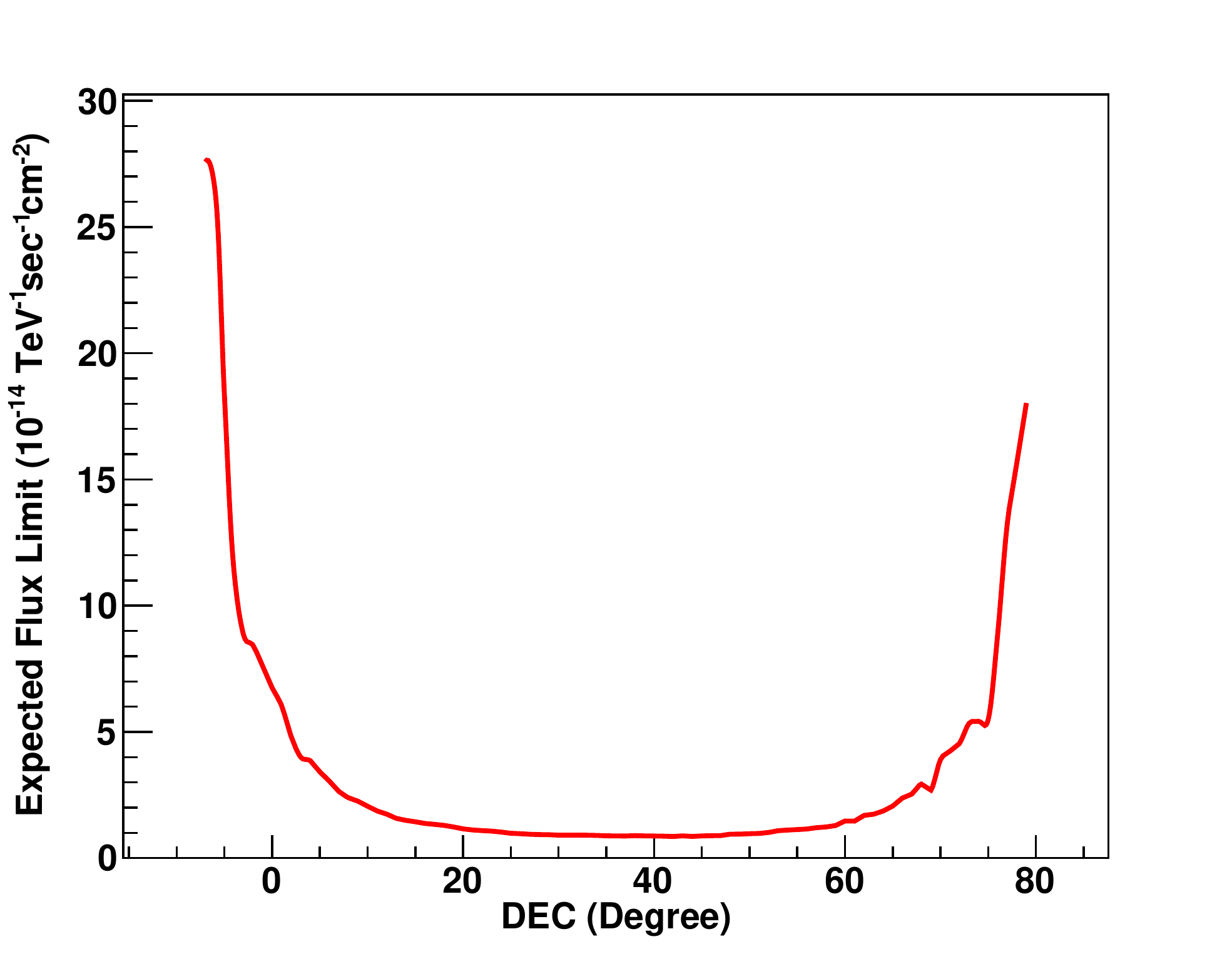}
\caption{The expected 95\% confidence level upper limit on the flux for extragalactic sources corresponding to zero excess derived at 7 TeV for each declination band of the Milagro sky maps made with spectral assumption $dN/dE \propto E^{-2.0}e^{-\frac{E}{5 TeV}}$.}\label{UpperLimit2p0}
\end{figure}

In this paper, we report the flux for the candidates with TeV associations that are identified by the FDR procedure. For the remaining
candidates we report flux upper limit. In all cases, the flux and significance calculations are performed assuming that the target is a point-like source.
The fluxes are calculated from the excess number of photons above background integrating over a Gaussian point spread function\footnote{The width of the Gaussian point spread function is a function of the estimated energy of each event and varies between $0.3^\circ$ and $0.7^\circ$.} for a point source at the sky position given by the catalog used to compile the list. This approach is similar to that described in the Milagro Galactic Plane Survey papers. The upper limits on the flux are determined using the method described in \cite{Helene} and are based on an upper limit on the number of excess photons with a 95\% confidence limit.
The flux upper limit corresponding to a zero excess is called the expected flux limit. The declination dependence of the expected flux limits shown in Figures ~\ref{UpperLimit2p6} and ~\ref{UpperLimit2p0} are based on Milagro maps made with the spectral optimizations $dN/dE \propto E^{-2.6}$ and $dN/dE \propto E^{-2.0}e^{-\frac{E}{5 TeV}}$ , respectively. The searches presented in this paper did not examine the whole sky. Another publication is in progress to produce all-sky flux limits from Milagro.\\

\section[]{Results} \label{results}

\begin{figure}
\centering
\includegraphics[width=50mm]{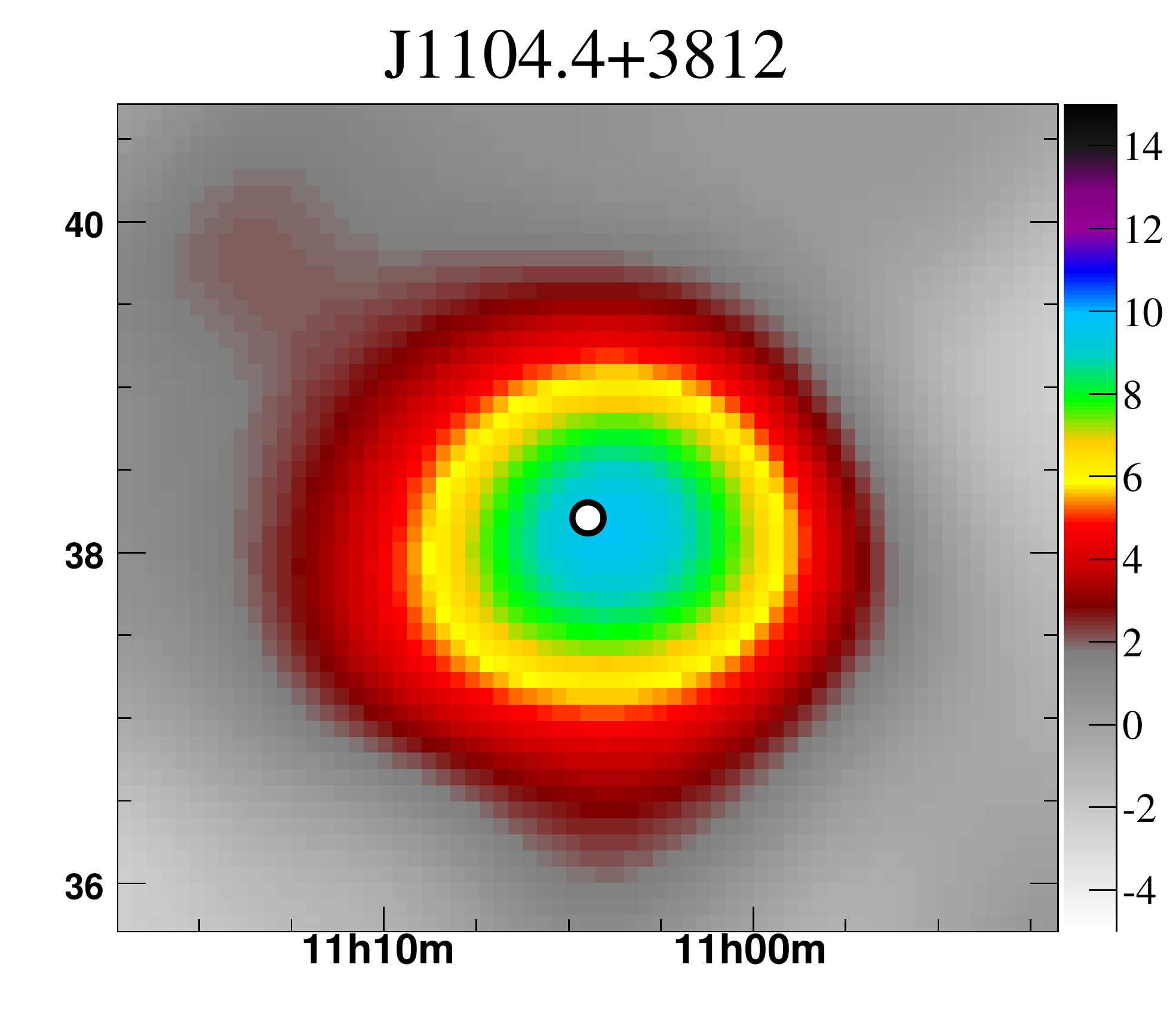}
\caption{This map shows the $5^\circ\times5^\circ$ (25 minutes $\times~5^\circ$) region around Mkn 421. 
The LAT source position is marked by a white dot. 
This map is made with the spectral optimization $dN/dE \propto E^{-2.0}e^{-\frac{E}{5TeV}}$  and the data have been smoothed using a Gaussian function.
The color of a bin shows the statistical significance (in standard deviations) of that bin. The horizontal axis is right ascension in hours and the vertical axis is declination in degrees.}\label{Mkn421}
\end{figure}

\begin{figure}
\centering
\subfigure[]{\includegraphics[width=50mm]{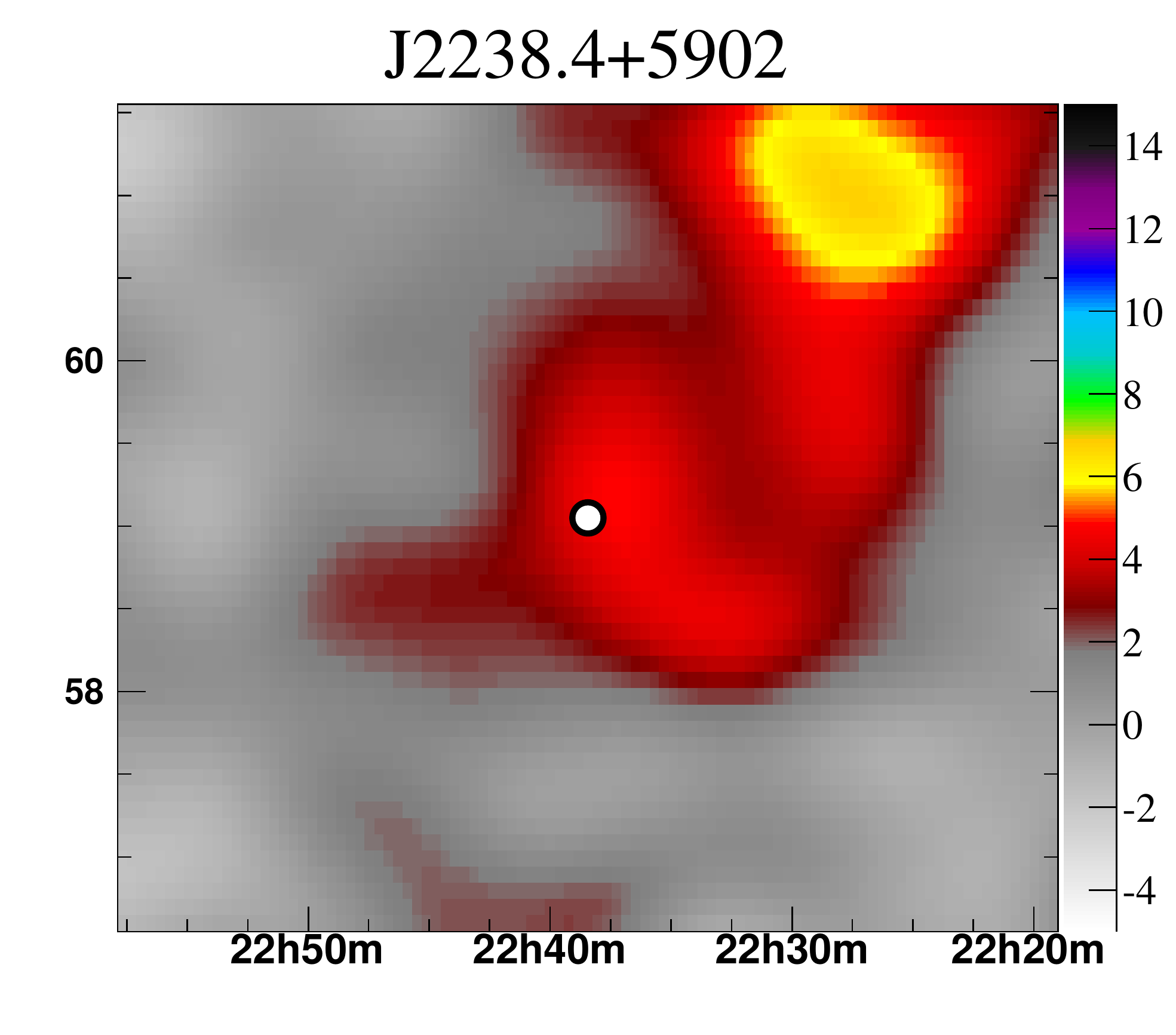}\label{2238}}
\subfigure[]{\includegraphics[width=50mm]{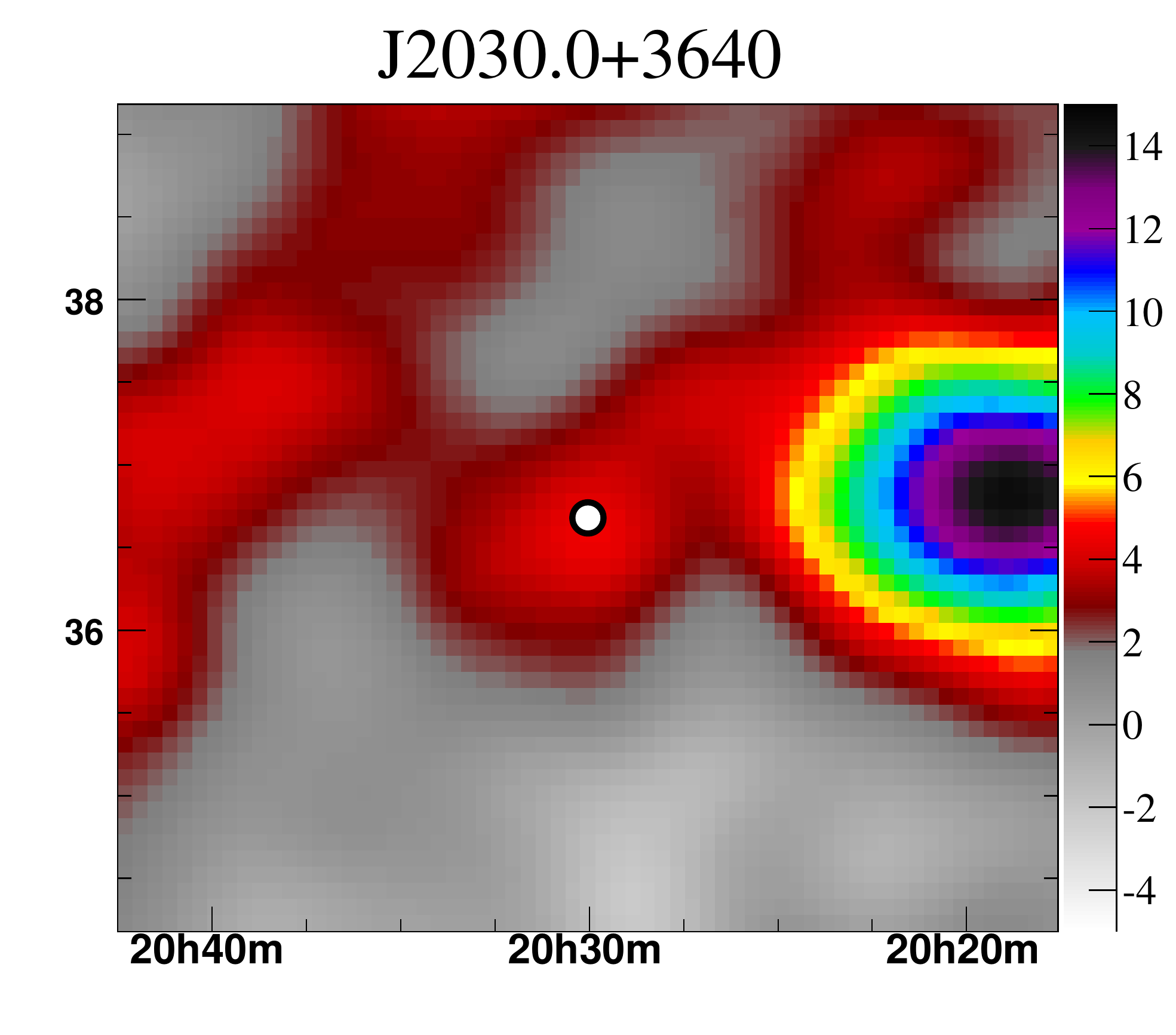}\label{2030}}
\subfigure[]{\includegraphics[width=50mm]{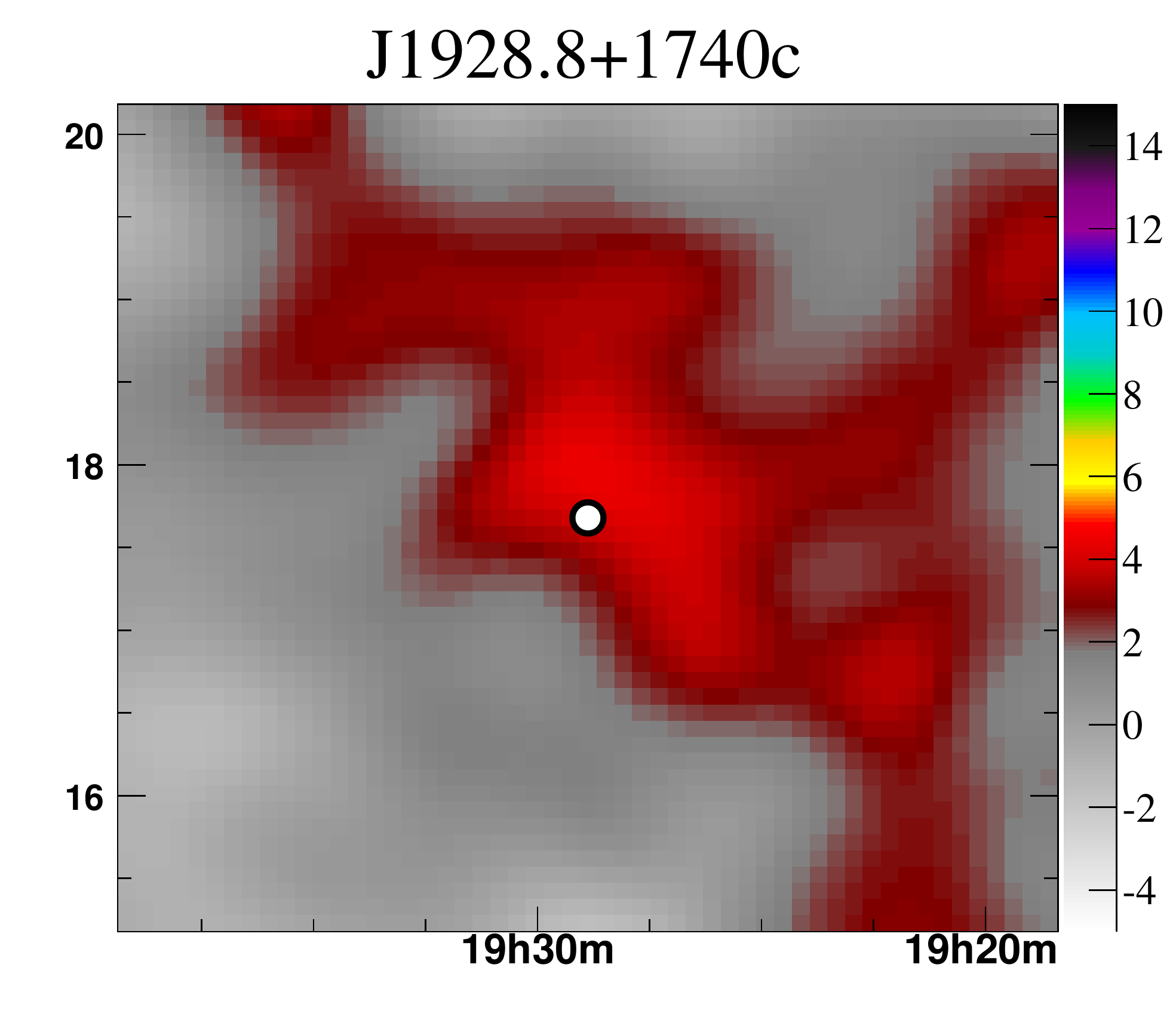}}
\caption{These maps show the $5^\circ\times5^\circ$ (25 minutes $\times~5^\circ$) region around 2FGL J2238.4+5902, 2FGL J2030.0+3640 and 2FGL J1928.8+1740c. The LAT source positions are marked by white dots. These maps are made with the spectral optimization $dN/dE \propto E^{-2.6}$  and the data have been smoothed using a Gaussian function.
The color of a bin shows the statistical significance (in standard deviations) of that bin.  The horizontal axis is right ascension in hours and the vertical axis is declination in degrees.}\label{PulsarSources}
\end{figure}

In the 2FGL Extragalactic List
only 2FGL J1104.4+3812 (also known as Mkn
421) is classified as a source by the FDR procedure with our
standard $\lambda=0.01$ cut. Figure~\ref{Mkn421} shows the region in the Milagro sky map around Mkn 421. From the 2FGL Extragalactic candidate list the fluxes or the 95\% confidence level flux upper limits for the brightest 20\% of the 2FGL candidates in the 3 GeV to 10 GeV energy band is given in Table~\ref {tab:FGL1}.
From the TeVCat Extragalactic List only Mkn 421 is classified as a source by the FDR procedure with our standard $\lambda=0.01$. The results with our standard FDR cut of $\lambda=0.01$ are summarized in Table~\ref{tab:TevCatSummary}.

Results from the source search in the Pulsar List are summarized in Table~\ref{tab:FGLPulsars}. 
In this list, the FDR procedure with $\lambda=0.01$ classified 3 GeV pulsars (2FGL J2238.4+5902, 2FGL J2030.0+3640 and 2FGL J1928.8+1740c) as having coincident TeV emission. Figure~\ref{PulsarSources} shows the regions of the Milagro sky maps around those candidates. 
The brighter area near 2FGL J2238.4+5902 (Figure~\ref{2238}) corresponds to 0FGL J2229.0+6114, and it is also associated with the bright TeV source MGRO J2228+61 \cite{MGRO2228+61}. 
The Milagro flux at the location of 0FGL J2229.0+6114 was published in the Milagro 0FGL paper.  
Similarly 2FGL J2030.0+3640 (Figure~\ref{2030}) is located near a brighter area which belongs to 0FGL J2020.8+3649. The Milagro flux at this 0FGL source location was also published in the Milagro
0FGL paper. 
Follow-up observations by TeV instruments with better angular resolution could clarify the TeV emission structure in both of these regions; some initial studies of this region have been already done [\cite{MilagroCygnus}, \cite{ARGOMilagroSources}]. 2FGL J2030.0+3640 also has a spatial association with the Milagro candidate named as C3 in \cite{Abdo2007}. Milagro candidate C3 is measured at RA = $307.75^0$ and DEC = $36.52^0$ with extent diameter of $2.8^0$ \citep{Abdo2007}. 

To assess how likely it would be to observe TeV emission coincident with 2FGL sources associated with pulsars if they arose from statistical fluctuations in the Milagro data, we calculate the probability that a set of 32 random points in the Milagro Galactic plane ($|l|<10^0$) would have 3 or more FDR associations. For a simulated background-only sky (consisting of a standard normal significance distribution) the probability of finding 3 or more associations is $1\times 10^{-6}$. Thus finding 3 associations would be a 4.3 $\sigma$ fluctuation for random points on a  background-only sky.  As expected, the $\lambda =0.01$ FDR cut yields no associations 99\% of the time with random locations on a random sky (with no real sources). However, the probability of finding 3 or more associations from random lists of 32 locations within the actual Milagro Galactic plane $|l|<10^0$ (which contains TeV sources) is 0.01. This is much higher than for a background-only sky, so that reporting 3 associations in the Milagro Galactic plane data would be a 2.3 $\sigma$ fluctuation if we were starting from a randomly located candidate list (rather than seeking associations with the  2FGL pulsar list).\\

By varying $\lambda$ the reader can construct alternative target lists with different contamination fractions, to assess how clearly candidates have passed a given association criterion. Table~\ref{tab:Summary} summarizes the FDR significance thresholds for each of the lists we have searched using $\lambda=0.1,0.05,0.01$ and $0.001$. The table also gives the significance thresholds which would have resulted from the trials correction technique. The comparison between the FDR and the trials corrections thresholds allows assessment of how much the FDR procedure has lowered the significance threshold in response to evidence of associations.\\

So far our results have focused on individual candidates with a TeV association. We also searched for evidence of collective TeV emission on the candidates that fail the $\lambda=0.01$ FDR cut by using the stacking method described in section~\ref{sec:StackingMethodology} . We stacked 2FGL Extragalactic candidates in two different ways: first all FDR False 2FGL Extragalactic sources and then the FDR False sources among the brightest 20\% in the \textit{Fermi-LAT} energy band 3-10 GeV. These two lists had $0.7 \sigma$ and $0.6 \sigma$ significance collectively. Stacking of TeV Cat candidates other than Mkn 421 has only a slightly more positive upward fluctuation of $0.9 \sigma$ . The rejected pulsar candidates have a $-0.5 \sigma$ fluctuation from the background. None of these stacking results indicate significant collective gamma-ray emission from the rejected candidates.

\section[]{Discussion}\label{discussion}

Mkn 421 is the only source that is classified as a TeV source in both extragalactic lists. Milagro  also observed a signal excess at the sky locations of Mkn 501, TXS 1720+102  and 1ES 0502+675 . Their significances are 2.93, 2.84 and 2.53 respectively, which is insufficient to pass our standard FDR cut of $\lambda=0.01$. Among these three candidates Mkn 501 and 1ES 0502+675 have been already reported as TeV sources in  TevCat. However TXS 1720+102 has not yet been identified as a source with TeV emission. This is a radio quasar type blazar identified at a redshift of 0.732 \citep{TXSRedshift}. The lowest FDR cut that TXS 1720+102 passes is $\lambda=0.32$.  With this loose FDR cut, three candidates become TeV associations: Mkn 421, Mkn 501 and TXS 1720+102. However, the expected contamination of the resulting  candidates list is 32\% so it is likely that TXS 1720+102 is a background fluctuation. While it is hard to advocate a dedicated IACT observation of TXS 1720+102, better observations will be performed by the High Altitude Water Cherenkov (HAWC) survey instrument \citep{HAWCRef}, which is already started to operate at a sensitivity better than Milagro.\\

In this paper we presented the TeV flux/flux limit measurements at 32 sky locations of 2FGL sources marked as pulsars.
At the time we wrote this paper, none of these 2FGL pulsars were reported as detections in the TeVCat or in the H.E.S.S. source catalog.
However, TeV flux upper limits of some these sources have been measured by IACTs.
For an example, the flux upper limit of PSR J1928+1746 was measured by the VERITAS observatory \cite{VERITASJ1928}, which is associated with the 2FGL J1928.8+1740c.
VERITAS observed a $+1.2\sigma$ significance at this pulsar position and 99\% confidence flux upper limit of $2.6\times10^{-13}$ cm$^{-2}$s$^{-1}$ above 1 TeV was measured assuming a power-law spectrum with power law index -2.5.
Contrasted with this measurement, Milagro measured a $46.41\pm11.5\times10^{-17}$ photons TeV$^{-1}$s$^{-1}$cm$^{-2}$ of flux at 35 TeV from this pulsar position, assuming a power-law spectrum with power law index -2.6.
The Milagro flux measurement is order of magnitude larger than the VERITAS upper limit.
This difference may be caused by the wider point spread function of Milagro compared with that of VERITAS ($\sim0.11^\circ$\cite{VERITASJ1928}).
Therefore, the Milagro flux may include some additional diffuse emission or emission from unresolved point sources that is not contained within the VERITAS point spread function.
We also compared our flux/flux limit measurements with the H.E.S.S. Galactic Plane Survey \cite{HESSGalacticSurvey}, and found that our measurements are consistent with the H.E.S.S. measurements.

The Milagro 0FGL paper reported the Milagro flux/flux limit at the locations of 16 bright \textit{Fermi-LAT} sources from the 0FGL catalog that were associated with pulsars.
Among these 16 pulsars, 9 passed the standard FDR cut of $\lambda=0.01$.
0FGL J2055.5+2540, 0FGL J2214.8+3002 and 0FGL J2302.9+443 were categorized as sources with unknown source type and 0FGL J1954.4+2838 was identified as a source with a spatial association with a known supernova remnant. 
In the 2FGL catalog these four sources have been identified as pulsars and only 0FGL J1954.4+2838 passed our standard FDR cut.
Therefore all together 52 pulsars detected by \textit{Fermi-LAT} have been observed by Milagro and 13 pulsars were identified with TeV associations.
We use this sample to study the correlation between GeV and TeV flux.\\

\begin{figure}
\centering
\includegraphics[width=\textwidth]{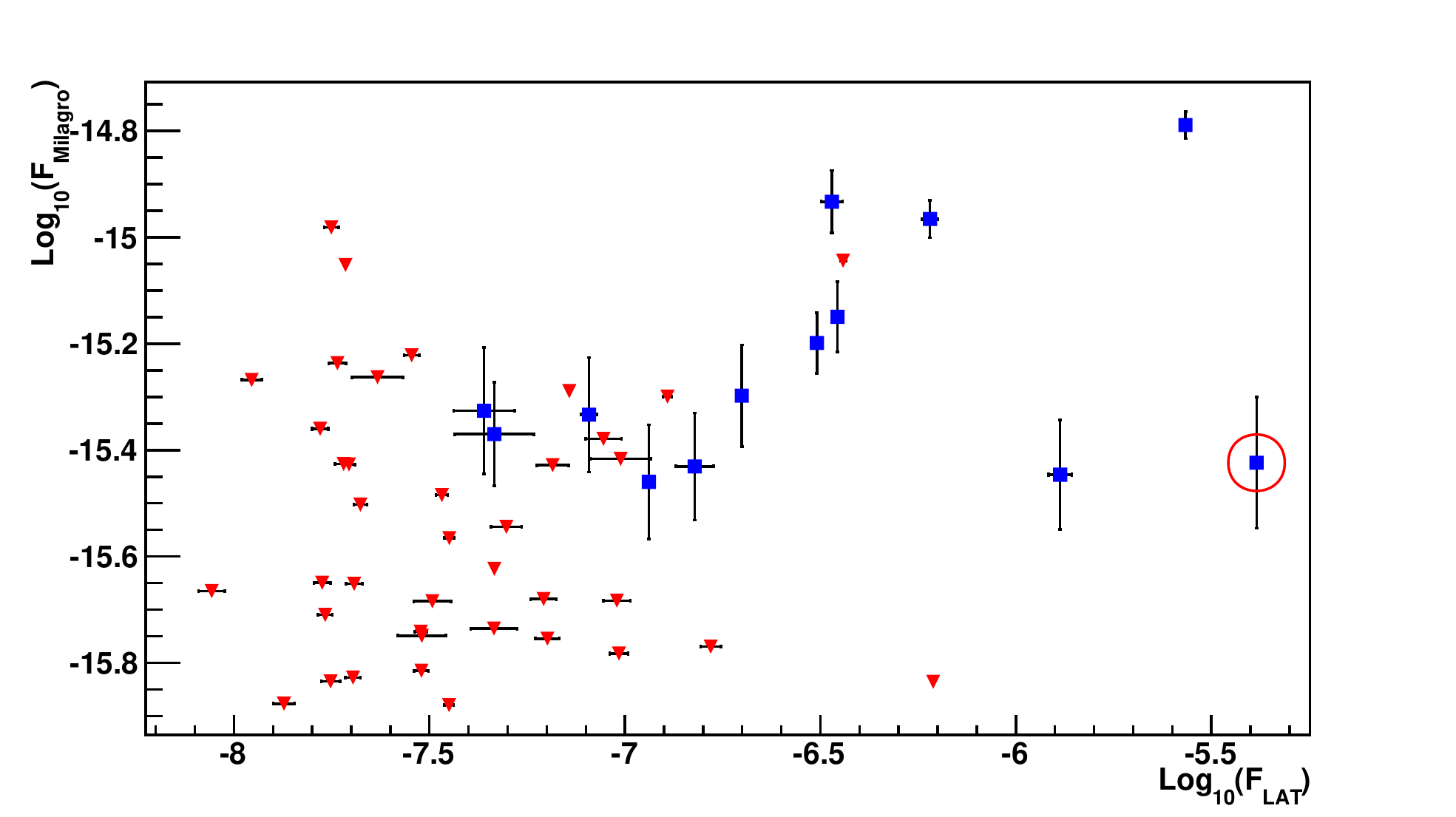}
\caption{The horizontal axis is the \textit{Fermi-LAT} flux (photons s$^{-1}$cm$^{-2}$), integrated over the energy range from 100 MeV to 100 GeV. The vertical axis is the Milagro flux derived at 35 TeV ( photons TeV$^{-1}$s$^{-1}$cm$^{-2}$), assuming all targets are point sources. Red data points are Milagro upper limits of candidates that failed the $\lambda=0.01$ FDR cut. Blue data points are the Milagro flux derived for the candidates that passed the $\lambda=0.01$ FDR cut. 
The Milagro flux/flux limits used in this plot were derived assuming the targets are point sources.
However, some of these objects are extended sources, for which the point source flux would underestimate the total flux. 
Geminga is a specific example and it is circled in red.}\label{Correlation}
\end{figure}

Figure~\ref{Correlation} shows the TeV flux measured by Milagro vs the GeV flux measured by \textit{Fermi-LAT} for these 52 pulsars. 
Data points marked with red triangles are Milagro upper limits measured at the sky locations of the candidates that failed the $\lambda=0.01$ FDR cut. 
Blue data points represents the Milagro flux at the sky locations of the candidates that passed the $\lambda=0.01$ FDR cut. 
The Milagro flux/flux limits used in this plot were derived assuming the targets are point sources. However, some of these objects are extended sources, for which the point source flux would underestimate the total flux. Geminga is a specific example, as seen in the Milagro 0FGL paper. In Figure~\ref{Correlation} Geminga is circled in red.\\

\begin{figure}
\centering
\includegraphics[width=100mm]{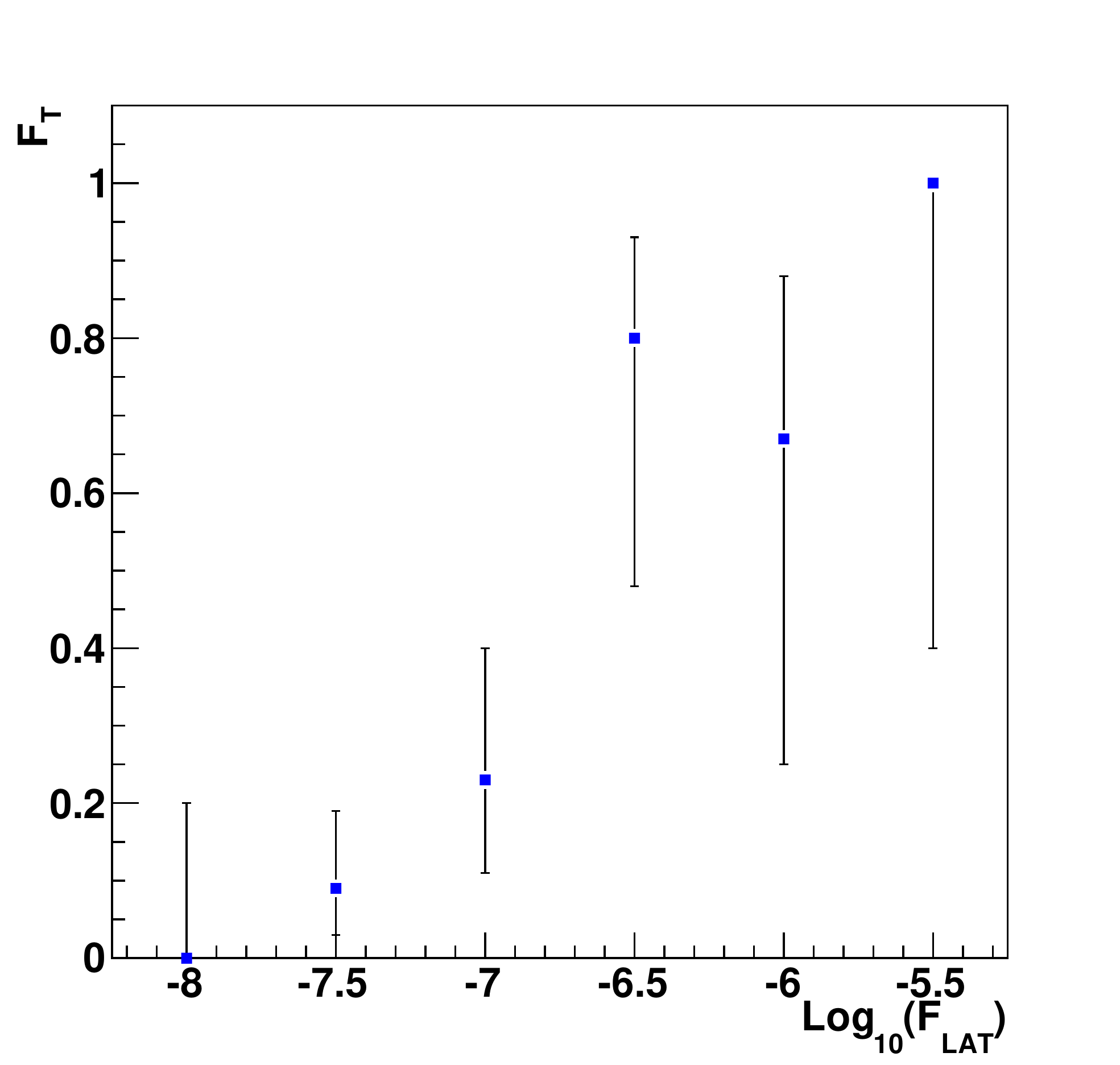}
\caption{The fraction $F_T$ (see text) of \textit{Fermi-LAT} pulsars seen by Milagro as a function of half-decade bins of the integrated \textit{Fermi-LAT} flux ( photons $cm^{-2}s^{-1}$) in the energy range from 100 MeV to 100 GeV.
}\label{efficiency}
\end{figure}

We can also study how the fraction of pulsars with a TeV counterpart changes as a function of the GeV flux.  
We define   $F_T$ as the fraction of pulsars that passed our standard FDR cut in a given bin of GeV flux.
\begin{equation}\label{PTDeff}
F_T = \frac{ \textnormal{Number of FDR true candidates in a given flux bin} }{ \textnormal{ Total number of candidates in a given flux bin} }
\end{equation}
As shown in Figure~\ref{efficiency} $F_T$ clearly increases with the \textit{Fermi-LAT} flux. Both the $F_T$ plot and the flux correlation plot strongly prefer a dependence on the GeV flux.  Therefore we have evidence that pulsars brighter in the GeV energy range are more likely to have a detectable TeV counterpart than pulsars fainter in the GeV energy range.  Further analysis of the GeV-TeV correlation is in progress and will be published in a follow-up paper.\\

\section[]{Conclusions}\label{Conclusion}
We present a targeted search for extragalactic sources in the Milagro data using a list of bright 2FGL extragalactic sources and TeV sources from the TeVCat catalog as targets.
Using the FDR procedure with $\lambda = 0.01$, we find that Mkn 421 is the only extragalactic TeV source detected by Milagro.
There is no evidence of collective TeV emission seen from the remaining extragalactic candidates.\\

The analysis performed in the Milagro 0FGL paper has been extended by searching for TeV emission at the locations of 32 additional \textit{Fermi-LAT} detected pulsars. TeV emission has been found associated with three of them:  2FGL J2238.4+5902, 2FGL J2030.0+3640 and 2FGL J1928.8+1740c. 
The first two of these are near bright VHE sources previously reported as being associated with energetic pulsars in the 0FGL catalog.
They might benefit from a higher spatial resolution TeV follow-up to study the emission structure from the two nearby source regions.  The pulsar candidates that failed the $\lambda = 0.01$ FDR cuts were studied in a stacking analysis but did not show any collective TeV emission. Finally, we presented evidence that pulsars brighter in the GeV energy range are more likely to have a detectable TeV counterpart.\\

\begin{table*}
\scriptsize
\centering
  \caption{Summary of the search with $\lambda = 0.01$ for TeV emission from the 2FGL list that  are identified as candidates off the galactic plane. (Note that we used the same abbreviations for the source type as the 2FGL, agu = active galaxy of uncertain type, bzb = BL Lac type of blazar and bzq = FSRQ type of blazar.) The Milagro flux derived at 7 TeV is given for candidates that passed the $\lambda = 0.01$ FDR cut and 95\% confidence level flux upper limit is given for the rest.\label{tab:FGL1}}
  \begin{tabular}{@{}llcccccccc}
  \hline
  \hline
  Fermi Name & RA & DEC &  l  &  b  & Flux/Flux Limit  &  Source Type  & Significance & Associated source\\
    2FGL        &  (deg)   &(deg)&  (deg)  &    (deg)   &      ($\times 10^{-17} \mbox{TeV}^{-1}$                       &                         & $\sigma$        &                    \\
            &       &   &       &       &   $\mbox{s}^{-1}\mbox{cm}^{-2}$)                  &           &           &           \\
\hline \hline
J0007.8+4713 & 1.97 & 47.23 & 115.3 & -15 & $<\ $65.06 & bzb & -0.78 &  MG4 J000800+4712 \\								
J0009.1+5030 & 2.29 & 50.51 & 116.09 & -11.8 & $<\ $85.8 & agu & -0.26 &  NVSS J000922+503028 \\								
J0022.5+0607 & 5.64 & 6.12 & 110.02 & -56.02 & $<\ $279.67 & bzb & -0.1 &  PKS 0019+058 \\								
J0045.3+2127 & 11.34 & 21.45 & 121.04 & -41.4 & $<\ $120.71 & bzb & 0.48 &  GB6 J0045+2127 \\								
J0100.2+0746 & 15.06 & 7.78 & 126.74 & -55.03 & $<\ $378.43 & bzb & 1.69 &  GB6 J0100+0745 \\								
\\
J0106.5+4854 & 16.65 & 48.91 & 125.49 & -13.88 & $<\ $95.99 & $\ $ & 0.21 &  $\ $ \\								
J0108.6+0135 & 17.17 & 1.59 & 131.85 & -60.98 & $<\ $566.7 & bzq & 0.64 &  4C 1.02 \\								
J0112.1+2245 & 18.03 & 22.76 & 129.15 & -39.86 & $<\ $63.03 & bzb & -1.29 &  S2 0109+22 \\								
J0112.8+3208 & 18.21 & 32.14 & 128.19 & -30.51 & $<\ $48.2 & bzq & -1.71 &  4C 31.03 \\								
J0115.4+0358 & 18.87 & 3.97 & 134.43 & -58.37 & $<\ $296.62 & bzb & -0.59 &  PMN J0115+0356 \\								
\\
J0136.5+3905 & 24.14 & 39.09 & 132.42 & -22.95 & $<\ $70.3 & bzb & -0.46 &  B3 0133+388 \\								
J0136.9+4751 & 24.24 & 47.86 & 130.78 & -14.32 & $<\ $117.84 & bzq & 0.89 &  OC 457 \\								
J0144.6+2704 & 26.16 & 27.08 & 137.29 & -34.31 & $<\ $85.04 & bzb & -0.03 &  TXS 0141+268 \\								
J0153.9+0823 & 28.49 & 8.4 & 148.21 & -51.38 & $<\ $209.19 & bzb & -0.2 &  GB6 J0154+0823 \\								
J0211.2+1050 & 32.81 & 10.84 & 152.59 & -47.39 & $<\ $106.04 & bzb & -1.48 &  MG1 J021114+1051 \\								
\\
J0217.4+0836 & 34.35 & 8.61 & 156.17 & -48.63 & $<\ $138.64 & bzb & -1.14 &  ZS 0214+083 \\								
J0217.9+0143 & 34.48 & 1.73 & 162.2 & -54.41 & $<\ $347.2 & bzq & -0.84 &  PKS 0215+015 \\								
J0221.0+3555 & 35.27 & 35.93 & 142.6 & -23.49 & $<\ $85.82 & bzq & 0.21 &  S4 0218+35 \\								
J0222.6+4302 & 35.66 & 43.04 & 140.14 & -16.77 & $<\ $85.18 & BZB & 0.13 &  3C 66A \\								
J0237.8+2846 & 39.47 & 28.78 & 149.48 & -28.55 & $<\ $101.74 & bzq & 0.58 &  4C 28.07 \\								
\\
J0238.7+1637 & 39.68 & 16.62 & 156.78 & -39.1 & $<\ $131.21 & BZB & 0.24 &  AO 0235+164 \\								
J0316.1+0904 & 49.05 & 9.08 & 172.1 & -39.59 & $<\ $339.0 & bzb & 1.59 &  GB6 J0316+0904 \\								
J0319.8+4130 & 49.97 & 41.51 & 150.58 & -13.25 & $<\ $74.54 & rdg & -0.15 &  NGC 1275 \\								
J0326.1+0224 & 51.55 & 2.41 & 180.74 & -42.45 & $<\ $802.03 & bzb & 1.66 &  1H 0323+022 \\								
J0333.7+2918 & 53.43 & 29.31 & 160.49 & -21.49 & $<\ $103.39 & agu & 0.62 &  TXS 0330+291 \\								
\\
J0423.2-0120 & 65.81 & -1.34 & 195.28 & -33.15 & $<\ $1039.11 & BZQ & 1.35 &  PKS 0420-01 \\								
J0433.5+2905 & 68.39 & 29.09 & 170.52 & -12.62 & $<\ $114.07 & bzb & 0.93 &  MG2 J043337+2905 \\								
J0442.7-0017 & 70.69 & 0.29 & 197.21 & -28.44 & $<\ $863.75 & bzq & 1.04 &  PKS 0440-00 \\								
J0448.9+1121 & 72.24 & 11.36 & 187.4 & -20.77 & $<\ $198.53 & bzq & 0.41 &  PKS 0446+11 \\								
J0508.0+6737 & 77.01 & 67.63 & 143.8 & 15.9 & $<\ $648.35 & bzb & 2.53 &  1ES 0502+675 \\								
\\
J0509.4+0542 & 77.37 & 5.7 & 195.4 & -19.62 & $<\ $320.94 & bzb & 0.34 &  TXS 0506+056 \\								
J0532.7+0733 & 83.19 & 7.56 & 196.84 & -13.71 & $<\ $275.2 & bzq & 0.68 &  OG 50 \\								
J0534.8-0548c & 83.72 & -5.81 & 209.36 & -19.66 & $<\ $942.7 & $\ $ & -1.51 &  $\ $ \\								
J0541.8-0203c & 85.45 & -2.06 & 206.69 & -16.41 & $<\ $1007.43 & $\ $ & 0.66 &  $\ $ \\								
J0547.1+0020c & 86.8 & 0.34 & 205.15 & -14.1 & $<\ $666.58 & $\ $ & 0.24 &  $\ $ \\								
\\
J0607.4+4739 & 91.87 & 47.66 & 165.64 & 12.87 & $<\ $76.14 & bzb & -0.4 &  TXS 0603+476 \\								
J0612.8+4122 & 93.21 & 41.37 & 171.83 & 10.92 & $<\ $86.2 & bzb & 0.24 &  B3 0609+413 \\								
J0650.7+2505 & 102.7 & 25.1 & 190.24 & 11.02 & $<\ $45.99 & bzb & -2.17 &  1ES 0647+250 \\								
J0654.2+4514 & 103.57 & 45.24 & 171.2 & 19.36 & $<\ $77.21 & bzq & -0.14 &  B3 0650+453 \\								
J0654.5+5043 & 103.65 & 50.72 & 165.68 & 21.14 & $<\ $164.55 & bzq & 1.79 &  GB6 J0654+5042 \\								
\\
J0714.0+1933 & 108.51 & 19.57 & 197.68 & 13.61 & $<\ $103.44 & bzq & -0.03 &  MG2 J071354+1934 \\								
J0719.3+3306 & 109.83 & 33.11 & 185.06 & 19.85 & $<\ $115.37 & bzq & 1.02 &  B2 0716+33 \\								
J0721.9+7120 & 110.48 & 71.35 & 143.97 & 28.02 & $<\ $512.3 & bzb & 0.49 &  S5 0716+71 \\								
J0725.3+1426 & 111.33 & 14.44 & 203.63 & 13.93 & $<\ $169.64 & BZQ & 0.49 &  4C 14.23 \\								
J0738.0+1742 & 114.52 & 17.7 & 201.85 & 18.06 & $<\ $152.56 & bzb & 0.83 &  PKS 0735+17 \\															 
\hline
\end{tabular}
\end{table*}

\addtocounter{table}{-1}

\begin{table*}
\scriptsize
\centering

 \caption{Cont.. Summary of the search with $\lambda = 0.01$ for TeV emission from the 2FGL list that  are identified as candidates off the galactic plane. (Note that we used the same abbreviations for the source type as the 2FGL, agu = active galaxy of uncertain type, bzb = BL Lac type of blazar and bzq = FSRQ type of blazar.) The Milagro flux derived at 7 TeV is given for candidates that passed the $\lambda = 0.01$ FDR cut and 95\% confidence level flux upper limit is given for the rest.\label{tab:FGL2}}
  \begin{tabular}{@{}llcccccccc}
  \hline
  \hline
  Fermi Name & RA & DEC &  l  &  b  & Flux/Flux Limit  &  Source Type  & Significance & Associated source\\
    2FGL        &  (deg)   &(deg)&  (deg)  &    (deg)   &      ($\times 10^{-17} \mbox{TeV}^{-1}$                       &                         & $\sigma$        &                    \\
            &       &   &       &       &   $\mbox{s}^{-1}\mbox{cm}^{-2}$)                  &           &           &           \\
\hline \hline	
J0739.2+0138 & 114.82 & 1.65 & 216.96 & 11.39 & $<\ $293.61 & bzq & -1.35 &  PKS 0736+01 \\								
J0805.3+7535 & 121.34 & 75.59 & 138.88 & 30.79 & $<\ $1032.12 & bzb & -0.11 &  RX J0805.4+7534 \\								
J0807.1-0543 & 121.78 & -5.72 & 227 & 13.99 & $<\ $1883.54 & bzb & 0.61 &  PKS 0804-05 \\								
J0809.8+5218 & 122.46 & 52.31 & 166.26 & 32.91 & $<\ $136.9 & bzb & 1.07 &  0806+524 \\								
J0818.2+4223 & 124.57 & 42.4 & 178.21 & 33.41 & $<\ $113.39 & bzb & 1.15 &  S4 0814+42 \\								
\\
J0831.9+0429 & 127.99 & 4.49 & 220.72 & 24.36 & $<\ $417.64 & bzb & 0.4 &  PKS 0829+046 \\								
J0854.8+2005 & 133.71 & 20.1 & 206.83 & 35.83 & $<\ $93.21 & BZB & -0.35 &  OJ 287 \\								
J0905.6+1357 & 136.4 & 13.96 & 215.04 & 35.96 & $<\ $190.04 & bzb & 0.88 &  MG1 J090534+1358 \\								
J0909.1+0121 & 137.29 & 1.37 & 228.93 & 30.92 & $<\ $576.86 & bzq & 0.19 &  PKS 0906+01 \\								
J0909.7-0229 & 137.43 & -2.5 & 232.8 & 28.99 & $<\ $476.09 & bzq & -1.54 &  PKS 0907-023 \\								
\\
J0915.8+2932 & 138.96 & 29.54 & 196.67 & 42.93 & $<\ $134.29 & bzb & 1.58 &  B2 0912+29 \\								
J0920.9+4441 & 140.24 & 44.7 & 175.7 & 44.81 & $<\ $107.06 & bzq & 0.84 &  S4 0917+44 \\								
J0957.7+5522 & 149.43 & 55.38 & 158.59 & 47.94 & $<\ $86.48 & bzq & -0.63 &  4C 55.17 \\								
J1012.6+2440 & 153.17 & 24.68 & 207.74 & 54.36 & $<\ $77.99 & bzq & -0.37 &  MG2 J101241+2439 \\								
J1015.1+4925 & 153.79 & 49.43 & 165.53 & 52.73 & $<\ $124.22 & bzb & 1.04 &  1H 1013+498 \\								
\\
J1016.0+0513 & 154.01 & 5.23 & 236.51 & 47.04 & $<\ $412.37 & bzq & 0.74 &  TXS 1013+054 \\								
J1033.9+6050 & 158.48 & 60.84 & 147.8 & 49.13 & $<\ $136.0 & BZQ & -0.18 &  S4 1030+61 \\								
J1037.6+5712 & 159.42 & 57.21 & 151.77 & 51.77 & $<\ $89.82 & bzb & -0.82 &  GB6 J1037+5711 \\								
J1058.4+0133 & 164.61 & 1.57 & 251.5 & 52.77 & $<\ $451.8 & bzb & 0.01 &  4C 1.28 \\								
J1058.6+5628 & 164.67 & 56.48 & 149.57 & 54.42 & $<\ $110.53 & bzb & 0.02 &  TXS 1055+567 \\								
\\
\textbf{J1104.4+3812} & \textbf{166.12} & \textbf{38.21} & \textbf{179.82} & \textbf{65.03} & \textbf{389.74$\pm $40.7} & \textbf{bzb} & \textbf{9.57} & \textbf{Mkn 421} \\								
J1117.2+2013 & 169.31 & 20.23 & 225.63 & 67.39 & $<\ $90.1 & bzb & -0.45 &  RBS 958 \\								
J1121.5-0554 & 170.39 & -5.91 & 266.27 & 50.45 & $<\ $2362.34 & bzq & 1.39 &  PKS 1118-05 \\								
J1132.9+0033 & 173.23 & 0.56 & 264.33 & 57.42 & $<\ $798.62 & bzb & 1.3 &  PKS B1130+008 \\								
J1150.5+4154 & 177.63 & 41.91 & 159.14 & 70.67 & $<\ $80.64 & bzb & 0.12 &  RBS 1040 \\								
\\
J1159.5+2914 & 179.88 & 29.25 & 199.41 & 78.37 & $<\ $79.56 & bzq & -0.11 &  Ton 599 \\								
J1217.8+3006 & 184.47 & 30.11 & 188.93 & 82.06 & $<\ $74.31 & bzb & -0.25 &  1ES 1215+303 \\								
J1221.3+3010 & 185.35 & 30.18 & 186.33 & 82.74 & $<\ $77.42 & bzb & -0.12 &  PG 1218+304 \\								
J1221.4+2814 & 185.37 & 28.24 & 201.69 & 83.28 & $<\ $113.67 & bzb & 0.9 &  W Comae \\								
J1224.9+2122 & 186.23 & 21.38 & 255.07 & 81.66 & $<\ $168.32 & BZQ & 1.59 &  4C 21.35 \\								
\\
J1226.0+2953 & 186.52 & 29.9 & 185.02 & 83.78 & $<\ $60.44 & $\ $ & -0.87 &  $\ $ \\								
J1229.1+0202 & 187.28 & 2.04 & 289.95 & 64.35 & $<\ $459.81 & BZQ & 0.01 &  3C 273 \\								
J1231.7+2848 & 187.94 & 28.81 & 190.66 & 85.34 & $<\ $54.27 & bzb & -1.28 &  B2 1229+29 \\								
J1239.5+0443 & 189.88 & 4.73 & 295.18 & 67.42 & $<\ $405.6 & bzq & 0.73 &  MG1 J123931+0443 \\								
J1243.1+3627 & 190.78 & 36.45 & 133.13 & 80.51 & $<\ $67.63 & bzb & -0.45 &  Ton 116 \\								
\\
J1248.2+5820 & 192.06 & 58.35 & 123.74 & 58.77 & $<\ $107.89 & bzb & -0.44 &  PG 1246+586 \\								
J1253.1+5302 & 193.28 & 53.05 & 122.36 & 64.08 & $<\ $72.69 & bzb & -0.99 &  S4 1250+53 \\								
J1256.1-0547 & 194.04 & -5.79 & 305.1 & 57.06 & $<\ $1072.78 & BZQ & -1.07 &  3C 279 \\								
J1303.1+2435 & 195.78 & 24.6 & 349.62 & 86.35 & $<\ $117.62 & bzb & 0.85 &  MG2 J130304+2434 \\								
J1309.4+4304 & 197.37 & 43.08 & 111.17 & 73.64 & $<\ $87.93 & bzb & 0.26 &  B3 1307+433 \\								
\\
J1310.6+3222 & 197.67 & 32.38 & 85.59 & 83.29 & $<\ $65.29 & bzq & -0.68 &  OP 313 \\								
J1312.8+4828 & 198.21 & 48.47 & 113.32 & 68.25 & $<\ $71.38 & bzq & -0.57 &  GB 1310+487 \\								
J1418.4-0234 & 214.6 & -2.57 & 341.56 & 53.64 & $<\ $1091.11 & bzb & 1.01 &  BZB J1418-0233 \\								
J1427.0+2347 & 216.76 & 23.8 & 29.48 & 68.2 & $<\ $96.03 & bzb & 0.13 &  PKS 1424+240 \\								
J1438.7+3712 & 219.68 & 37.21 & 63.72 & 65.27 & $<\ $50.77 & bzq & -1.33 &  B2 1436+37B \\								
																																	
\hline
 \end{tabular}
 
 \end{table*}

\addtocounter{table}{-1}

\begin{table*}
\scriptsize
\centering

 \caption{Cont.. Summary of the search with $\lambda = 0.01$ for TeV emission from the 2FGL list that  are identified as candidates off the galactic plane. (Note that we used the same abbreviations for the source type as the 2FGL, agu = active galaxy of uncertain type, bzb = BL Lac type of blazar and bzq = FSRQ type of blazar.) The Milagro flux derived at 7 TeV is given for candidates that passed the $\lambda = 0.01$ FDR cut and 95\% confidence level flux upper limit is given for the rest.\label{tab:FGL3}}
  \begin{tabular}{@{}llcccccccc}
  \hline
  \hline
  Fermi Name & RA & DEC &  l  &  b  & Flux/Flux Limit  &  Source Type  & Significance & Associated source\\
    2FGL        &  (deg)   &(deg)&  (deg)  &    (deg)   &      ($\times 10^{-17} \mbox{TeV}^{-1}$                       &                         & $\sigma$        &                    \\
            &       &   &       &       &   $\mbox{s}^{-1}\mbox{cm}^{-2}$)                  &           &           &           \\
\hline \hline											
J1454.4+5123 & 223.62 & 51.4 & 87.66 & 56.46 & $<\ $109.12 & bzb & 0.49 &  TXS 1452+516 \\								
J1501.0+2238 & 225.28 & 22.64 & 31.46 & 60.34 & $<\ $114.2 & bzb & 0.55 &  MS 1458.8+2249 \\								
J1504.3+1029 & 226.1 & 10.49 & 11.37 & 54.58 & $<\ $185.18 & BZQ & 0.0 &  PKS 1502+106 \\								
J1520.8-0349 & 230.22 & -3.83 & 358.11 & 42.48 & $<\ $820.68 & bzb & -0.48 &  NVSS J152048-034850 \\								
J1522.1+3144 & 230.54 & 31.74 & 50.18 & 57.02 & $<\ $101.94 & bzq & 0.64 &  B2 1520+31 \\								
\\
J1542.9+6129 & 235.73 & 61.49 & 95.38 & 45.4 & $<\ $93.21 & bzb & -1.34 &  GB6 J1542+6129 \\								
J1553.5+1255 & 238.39 & 12.93 & 23.77 & 45.21 & $<\ $109.28 & bzq & -0.97 &  PKS 1551+130 \\								
J1555.7+1111 & 238.94 & 11.19 & 21.92 & 43.95 & $<\ $160.64 & bzb & -0.17 &  PG 1553+113 \\								
J1607.0+1552 & 241.77 & 15.88 & 29.4 & 43.42 & $<\ $85.05 & bzb & -1.16 &  4C 15.54 \\								
J1625.2-0020 & 246.3 & 0.33 & 13.92 & 31.83 & $<\ $762.2 & $\ $ & 0.66 &  $\ $ \\								
\\
J1635.2+3810 & 248.81 & 38.17 & 61.13 & 42.34 & $<\ $58.66 & bzq & -0.95 &  4C 38.41 \\								
J1637.7+4714 & 249.43 & 47.24 & 73.38 & 41.88 & $<\ $68.76 & bzq & -0.6 &  4C 47.44 \\								
J1640.7+3945 & 250.18 & 39.76 & 63.35 & 41.38 & $<\ $122.41 & BZQ & 1.31 &  NRAO 512 \\								
J1642.9+3949 & 250.18 & 39.76 & 63.48 & 40.95 & $<\ $122.41 & BZQ & 1.31 &  3C 345 \\								
J1640.7+3945 & 250.75 & 39.83 & 63.35 & 41.38 & $<\ $115.23 & BZQ & 1.11 &  NRAO 512 \\								
\\
J1642.9+3949 & 250.75 & 39.83 & 63.48 & 40.95 & $<\ $115.23 & BZQ & 1.11 &  3C 345 \\								
J1653.6-0159 & 253.4 & -2 & 16.59 & 24.93 & $<\ $847.78 & $\ $ & 0.15 &  $\ $ \\								
J1653.9+3945 & 253.48 & 39.76 & 63.61 & 38.85 & $<\ $186.65 & BZB & 2.93 &  Mkn 501 \\								
J1700.2+6831 & 255.06 & 68.52 & 99.58 & 35.19 & $<\ $316.95 & bzq & -0.1 &  TXS 1700+685 \\								
J1709.7+4319 & 257.45 & 43.32 & 68.41 & 36.21 & $<\ $113.14 & bzq & 1.01 &  B3 1708+433 \\								
\\
J1719.3+1744 & 259.83 & 17.74 & 39.53 & 28.07 & $<\ $183.16 & bzb & 1.41 &  PKS 1717+177 \\								
J1722.7+1013 & 260.68 & 10.23 & 32.22 & 24.3 & $<\ $425.7 & bzq & 2.84 &  TXS 1720+102 \\								
J1725.0+1151 & 261.27 & 11.87 & 34.11 & 24.47 & $<\ $288.43 & bzb & 1.82 &  1H 1720+117 \\								
J1734.3+3858 & 263.58 & 38.98 & 64.04 & 31.02 & $<\ $95.96 & bzq & 0.48 &  B2 1732+38A \\								
J1748.8+7006 & 267.22 & 70.11 & 100.54 & 30.69 & $<\ $275.01 & bzb & -1.01 &  1749+70 \\								
\\
J1751.5+0938 & 267.88 & 9.64 & 34.91 & 17.65 & $<\ $183.43 & bzb & 0.0 &  OT 81 \\								
J1754.3+3212 & 268.58 & 32.2 & 57.75 & 25.38 & $<\ $60.01 & bzb & -0.97 &  RX J1754.1+3212 \\								
J1800.5+7829 & 270.15 & 78.48 & 110.06 & 29.07 & $<\ $1134.67 & bzb & -0.91 &  S5 1803+784 \\								
J1806.7+6948 & 271.68 & 69.8 & 100.1 & 29.18 & $<\ $497.66 & bzb & 0.75 &  3C 371 \\								
J1811.3+0339 & 272.83 & 3.66 & 31.62 & 10.59 & $<\ $279.71 & bzb & -0.73 &  NVSS J181118+034114 \\								
\\
J1824.0+5650 & 276 & 56.84 & 85.72 & 26.09 & $<\ $65.02 & bzb & -1.93 &  4C 56.27 \\								
J1838.7+4759 & 279.7 & 47.99 & 76.9 & 21.82 & $<\ $107.68 & bzb & 0.63 &  GB6 J1838+4802 \\								
J1849.4+6706 & 282.35 & 67.1 & 97.5 & 25.03 & $<\ $287.58 & bzq & 0.2 &  S4 1849+67 \\								
J1852.5+4856 & 283.13 & 48.94 & 78.6 & 19.94 & $<\ $41.8 & bzq & -2.49 &  S4 1851+48 \\								
J1903.3+5539 & 285.84 & 55.67 & 85.96 & 20.51 & $<\ $77.36 & bzb & -1.1 &  TXS 1902+556 \\								
\\
J1927.0+6153 & 291.77 & 61.9 & 93.31 & 19.71 & $<\ $123.93 & bzb & -0.79 &  1RXS J192649.5+615445 \\								
J2000.0+6509 & 300.02 & 65.16 & 98.02 & 17.67 & $<\ $208.54 & bzb & -0.07 &  1ES 1959+650 \\								
J2116.2+3339 & 319.05 & 33.66 & 79.82 & -10.64 & $<\ $124.07 & bzb & 1.32 &  B2 2114+33 \\								
J2121.0+1901 & 320.26 & 19.03 & 69.25 & -21.25 & $<\ $168.15 & bzq & 1.26 &  OX 131 \\								
J2133.9+6645 & 323.49 & 66.75 & 105.17 & 10.96 & $<\ $316.94 & $\ $ & 0.47 &  $\ $ \\								
\\
J2143.5+1743 & 325.88 & 17.72 & 72.09 & -26.08 & $<\ $108.01 & bzq & -0.21 &  OX 169 \\								
J2147.3+0930 & 326.84 & 9.51 & 65.85 & -32.28 & $<\ $162.46 & bzq & -0.35 &  PKS 2144+092 \\								
J2202.8+4216 & 330.71 & 42.27 & 92.6 & -10.46 & $<\ $47.82 & bzb & -1.53 &  BL Lacertae \\								
J2203.4+1726 & 330.87 & 17.44 & 75.68 & -29.63 & $<\ $164.9 & bzq & 0.9 &  PKS 2201+171 \\								
J2236.4+2828 & 339.1 & 28.48 & 90.12 & -25.66 & $<\ $86.69 & bzb & 0.05 &  B2 2234+28A \\																 
\hline
\end{tabular}

\end{table*}

\addtocounter{table}{-1}

\begin{table*}
\scriptsize
\centering

\caption{Cont.. Summary of the search with $\lambda = 0.01$ for TeV emission from the 2FGL list that  are identified as candidates off the galactic plane. (Note that we used the same abbreviations for the source type as the 2FGL, agu = active galaxy of uncertain type, bzb = BL Lac type of blazar and bzq = FSRQ type of blazar.) The Milagro flux derived at 7 TeV is given for candidates that passed the $\lambda = 0.01$ FDR cut and 95\% confidence level flux upper limit is given for the rest.\label{tab:FGL4}}
 \begin{tabular}{@{}llcccccccc}
 \hline
 \hline
 Fermi Name & RA & DEC &  l  &  b  & Flux/Flux Limit  &  Source Type  & Significance & Associated source\\
   2FGL        &  (deg)   &(deg)&  (deg)  &    (deg)   &      ($\times 10^{-17} \mbox{TeV}^{-1}$                       &                         & $\sigma$        &                    \\
           &       &   &       &       &   $\mbox{s}^{-1}\mbox{cm}^{-2}$)                  &           &           &           \\
\hline \hline														
J2243.9+2021 & 341 & 20.36 & 86.59 & -33.37 & $<\ $84.11 & bzb & -0.7 &  RGB J2243+203 \\								
J2244.1+4059 & 341.03 & 40.99 & 98.5 & -15.77 & $<\ $100.26 & bzb & 0.69 &  TXS 2241+406 \\								
J2253.9+1609 & 343.5 & 16.15 & 86.12 & -38.18 & $<\ $135.72 & BZQ & 0.23 &  3C 454.3 \\								
J2311.0+3425 & 347.77 & 34.43 & 100.42 & -24.02 & $<\ $62.49 & bzq & -0.78 &  B2 2308+34 \\								
J2323.6-0316 & 350.91 & -3.28 & 77.78 & -58.23 & $<\ $552.24 & bzq & -1.9 &  PKS 2320-035 \\								
\\
J2323.8+4212 & 350.95 & 42.2 & 106.06 & -17.78 & $<\ $81.89 & bzb & 0.13 &  1ES 2321+419 \\								
J2325.3+3957 & 351.33 & 39.96 & 105.52 & -19.98 & $<\ $47.82 & bzb & -1.62 &  B3 2322+396 \\								
J2334.8+1431 & 353.72 & 14.53 & 96.56 & -44.39 & $<\ $167.75 & bzb & 0.7 &  BZB J2334+1408 \\								
J2339.6-0532 & 354.91 & -5.54 & 81.36 & -62.47 & $<\ $2061.09 & $\ $ & 0.84 &  $\ $ \\					
\hline
\end{tabular}
\end{table*}

\begin{table*}
\scriptsize
\centering
  \caption{ Summary of the search with $\lambda = 0.01$ for TeV emission from the TeVCat list that are identified as candidates off the galactic plane. (Note that: HBL = High Frequency Peaked BL Lac, IBL = Intermediate Frequency Peaked BL Lac, LBL = Low Frequency Peaked BL Lac, UNID = Unidentified, FSRQ = Flat Spectrum Radio Quasar, AGN = Active Galactic Nuclei, Cat. Var. = Cataclysmic Variable Star and FR I = Fanaroff-Riley Type I radio source.) The Milagro flux derived at 7 TeV is given for the candidates that passes the $\lambda = 0.01$ FDR cut and 95\% confidence level upper limits given for the rest. \label{tab:TevCatSummary}}
  \begin{tabular}{@{}llcccccccc}
  \hline
  \hline
Name & RA       & DEC &  l  &  b  & Flux/Flux Limit  &  Source Type  & Significance & 2FGL Association \\
            &  (deg)   &(deg)&  (deg)  &    (deg)   &      ($\times 10^{-17} \mbox{TeV}^{-1}$                       &                         & $\sigma$        &      2FGL              \\
            &       &   &       &       &   $\mbox{s}^{-1}\mbox{cm}^{-2}$)                  &           &              &         \\
\hline \hline
 RGB J0152+017 & 28.1396 & 1.77786 & 152.34317 & -57.561295 & $<\ $467.36 & HBL & 0.04\\							
 3C66A & 35.6733 & 43.0432 & 140.24803 & -16.753392 & $<\ $85.18 & IBL & 0.13\\							
 3C66A/B & 35.8 & 43.0117 & 140.24803 & -16.753392 & $<\ $83.18 & UNID & 0.06\\							
 1ES 0229+200 & 38.2217 & 20.2725 & 152.97002 & -36.612512 & $<\ $81.55 & HBL & -0.8\\							
 IC 310 & 49.1792 & 41.3247 & 150.57567 & -13.261242 & $<\ $76.53 & AGN & -0.12\\							
\\
 NGC 1275 & 49.9504 & 41.5117 & 150.57567 & -13.261242 & $<\ $74.54 & FRI & -0.15\\							
 RBS 0413 & 49.9658 & 18.7594 & 165.10684 & -31.69731 & $<\ $97.2 & HBL & -0.4\\							
 1ES 0414+009 & 64.2184 & 1.09008 & 191.81416 & -33.159267 & $<\ $478.08 & HBL & -0.36\\							
 1ES 0502+675\footnote{This candidate passes the FDR cut $\lambda=0.1$.} & 76.9842 & 67.6233 & 143.795 & 15.88981 & $<\ $651.78 & HBL & 2.55\\							
 RGB J0710+591 & 107.61 & 59.15 & 157.39076 & 25.420975 & $<\ $146.09 & HBL & 0.31\\							
\\
 S5 0716+714 & 110.473 & 71.3433 & 143.9812 & 28.017623 & $<\ $512.3 & LBL & 0.49\\							
 1ES 0806+524 & 122.496 & 52.3167 & 166.24607 & 32.93548 & $<\ $136.9 & HBL & 1.07\\							
 M82 & 148.97 & 69.6794 & 141.4095 & 40.567564 & $<\ $343.13 & Starburst & -0.35\\							
 1ES 1011+496 & 153.767 & 49.4336 & 165.53394 & 52.712223 & $<\ $124.22 & HBL & 1.04\\							
\textbf{Markarian 421} & \textbf{166.079} & \textbf{38.1947} & \textbf{179.88395} & \textbf{65.01015} & \textbf{395.08$\pm $40.69} & \textbf{HBL} & \textbf{9.7}\\							
\\
 Markarian 180 & 174.11 & 70.1575 & 131.90989 & 45.641234 & $<\ $362.44 & HBL & -0.15\\							
 1ES 1215+303 & 184.467 & 30.1169 & 188.87483 & 82.052923 & $<\ $74.31 & LBL & -0.25\\							
 1ES 1218+304 & 185.36 & 30.1914 & 186.20601 & 82.743376 & $<\ $77.42 & HBL & -0.12\\							
 W Comae & 185.382 & 28.2331 & 201.735 & 83.288032 & $<\ $113.67 & IBL & 0.9\\							
 4C +21.35 & 186.227 & 21.3794 & 255.07319 & 81.65946 & $<\ $168.32 & FSRQ & 1.59\\							
\\
 M87 & 187.697 & 12.3975 & 283.73831 & 74.494439 & $<\ $115.55 & FRI & -1.02\\							
 3C279 & 194.046 & -4.21056 & 305.20657 & 58.640166 & $<\ $1212.26 & FSRQ & -0.11\\							
 PKS 1424+240 & 216.752 & 23.8 & 29.487026 & 68.207689 & $<\ $96.03 & IBL & 0.13\\							
 H 1426+428 & 217.136 & 42.6725 & 77.487039 & 64.899104 & $<\ $79.44 & HBL & -0.04\\							
 1ES 1440+122 & 220.701 & 12.0111 & 8.3294143 & 59.840034 & $<\ $96.45 & IBL & -1.62\\							
\\
 PG 1553+113 & 238.936 & 11.1947 & 21.918776 & 43.960313 & $<\ $160.64 & HBL & -0.17\\							
 Markarian 501\footnote{This candidate passes the FDR cut $\lambda=0.05$.}  & 253.468 & 39.7603 & 63.600083 & 38.859361 & $<\ $186.65 & HBL & 2.93\\							
 1ES 1959+650 & 299.999 & 65.1486 & 98.003397 & 17.670031 & $<\ $209.2 & HBL & -0.06\\							
 AEAquarii & 310.042 & 0.871111 & 46.934522 & -23.552194 & $<\ $446.02 & Cat.\_Var. & -0.53\\							
 BL Lacertae & 330.68 & 42.2778 & 92.589572 & -10.441029 & $<\ $48.15 & LBL & -1.51\\							
\\
 B3 2247+381 & 342.528 & 38.4328 & 98.267934 & -18.559532 & $<\ $84.69 & HBL & 0.14\\							

\hline
\end{tabular}
\end{table*}

\begin{table*}
\scriptsize
\centering
  \caption{Summary of the search with $\lambda = 0.01$ for TeV emission from the pulsars in the 2FGL list that were not listed in the 0FGL. (Note that we used the same abbreviations for the source type as the 2FGL:  PSR = Pulsar identified by pulsations and psr 	= Pulsar identifies by spatial association.) The Milagro flux derived at 35 TeV is given for the candidates that passed the $\lambda = 0.01$ FDR cut and 95\% confidence level upper limits are given for the rest. \label{tab:FGLPulsars}  }
  \begin{tabular}{@{}llccccccl}
  \hline
  \hline
  Fermi Name & RA & DEC &  l  &  b  & Flux/Flux Limit  & Source Type & Significance & Associated source\\
  2FGL		&  (deg)   &(deg)&  (deg)  &    (deg)   &      ($\times 10^{-17} \mbox{TeV}^{-1}$                       &                         & $\sigma$        &                    \\
  			&		&	&		&		&	$\mbox{s}^{-1}\mbox{cm}^{-2}$)					&			&			&		    \\
 \hline \hline																														
J0023.5+0924	&	5.89	&	9.41	&	111.5	&	-52.85	&	$<\	$22.42	&	psr	&	-0.73	&		PSR	J0023+09	\\										 
J0034.4-0534	&	8.61	&	-5.58	&	111.55	&	-68.08	&	$<\	$54.58	&	PSR	&	-2.06	&		PSR	J0034-0534	\\										 
J0102.9+4838	&	15.74	&	48.65	&	124.9	&	-14.18	&	$<\	$22.31	&	psr	&	0.51	&		PSR	J0103+48	\\										 
J0205.8+6448	&	31.45	&	64.81	&	130.74	&	3.07	&	$<\	$20.75	&	PSR	&	-0.88	&		PSR	J0205+6449	\\										 
J0218.1+4233	\footnote{This	candidate	passes the FDR cut	$\lambda=0.05$.}	&	34.53	&	42.55	&	139.5	&	-17.51	&	$<\	 $41.80	&	PSR	&	2.94	&		PSR	J0218+4232	\\
\\
J0248.1+6021	&	42.04	&	60.36	&	136.89	&	0.69	&	$<\	$38.32	&	PSR	&	1.54	&		PSR	J0248+6021	\\										 
J0308.3+7442	&	47.08	&	74.71	&	131.73	&	14.23	&	$<\	$53.95	&	psr	&	-0.15	&		PSR	J0308+7442	\\										 
J0340.4+4131	&	55.1	&	41.53	&	153.78	&	-11.01	&	$<\	$14.87	&	PSR	&	-0.52	&		PSR	J0340+4130	\\										 
J0659.7+1417	&	104.93	&	14.29	&	201.05	&	8.27	&	$<\	$37.31	&	PSR	&	1.18	&		PSR	J0659+1414	\\										 
J0751.1+1809	&	117.78	&	18.15	&	202.7	&	21.09	&	$<\	$19.51	&	PSR	&	-0.43	&		PSR	J0751+1807	\\										 
\\
J1023.6+0040	&	155.92	&	0.68	&	243.43	&	45.78	&	$<\	$43.66	&	psr	&	-0.33	&		PSR	J1023+0038	\\										 
J1142.9+0121	&	175.74	&	1.35	&	267.56	&	59.44	&	$<\	$37.41	&	psr	&	-0.9	&		PSR	J1142+01	\\										 
J1301.5+0835	&	195.39	&	8.58	&	310.76	&	71.3	&	$<\	$21.62	&	psr	&	-0.82	&		PSR	J1301+08	\\										 
J1312.7+0051	&	198.18	&	0.85	&	314.82	&	63.23	&	$<\	$58.01	&	psr	&	0.5	&		PSR	J1312+00	\\										 
J1549.7-0657	&	237.43	&	-6.96	&	1.23	&	35.03	&	$<\	$88.75	&	psr	&	-0.8	&		PSR	J1549-06	\\										 
\\
J1714.0+0751	&	258.5	&	7.86	&	28.84	&	25.21	&	$<\	$18.13	&	PSR	&	-1.58	&		PSR	J1713+0747	\\										 
J1745.6+1015	&	266.4	&	10.27	&	34.84	&	19.23	&	$<\	$32.77	&	psr	&	0.48	&		PSR	J1745+10	\\										 
J1810.7+1742	&	272.69	&	17.7	&	44.62	&	16.76	&	$<\	$18.39	&	psr	&	-0.6	&		PSR	J1810+17	\\										 
J1846.4+0920	&	281.61	&	9.34	&	40.7	&	5.34	&	$<\	$23.81	&	PSR	&	-0.54	&		PSR	J1846+0919	\\									 
\textbf{J1928.8+1740c}	& \textbf{292.22} & \textbf{17.68} & \textbf{52.87} & \textbf{0.03} & \textbf{46.41$\pm$11.50} &	\textbf{psr}  & \textbf{4.03}  & \textbf{PSR	J1928+1746}	\\							
\\		
J1957.9+5033	&	299.48	&	50.56	&	84.61	&	10.98	&	$<\	$28.56	&	PSR	&	1.41	&		PSR	J1957+5033	\\									 
J1959.5+2047	&	299.9	&	20.79	&	59.18	&	-4.7	&	$<\	$15.32	&	PSR	&	-0.85	&		PSR	J1959+2048	\\										 
\textbf{J2030.0+3640}	&	\textbf{307.51}	&	\textbf{36.68}	&	\textbf{76.12}	&	\textbf{-1.45}	&	\textbf{42.68$\pm$9.55} & \textbf{PSR} & \textbf{4.46} & \textbf{PSR J2030+3641}	\\	
J2017.3+0603 &  	304.35 	&  	6.05		&  	48.63	&	-16.02	&	$<\ $ 27.2	& PSR	&	-0.71	 &	PSR J2017+0603 \\								 
J2043.2+1711	&	310.81	&	17.18	&	61.9	&	-15.3	&	$<\	$17.82	&	PSR	&	-0.76	&		PSR	J2043+1710	\\										 
\\
J2043.7+2743	&	310.95	&	27.72	&	70.65	&	-9.14	&	$<\	$14.62	&	PSR	&	-0.72	&		PSR	J2043+2740	\\										 
J2046.7+1055	&	311.69	&	10.93	&	57.02	&	-19.57	&	$<\	$37.49	&	psr	&	0.99	&		PSR	J2047+10	\\										 
J2129.8-0428	&	322.47	&	-4.48	&	48.93	&	-36.96	&	$<\	$104.35	&	psr	&	0.55	&		PSR	J2129-04	\\										 
J2215.7+5135	&	333.94	&	51.59	&	99.89	&	-4.18	&	$<\	$13.28	&	psr	&	-1.1	&		PSR	J2215+51	\\										 
J2234.7+0945	&	338.69	&	9.75	&	76.29	&	-40.43	&	$<\	$31.44	&	psr	&	0.38	&		PSR	J2234+09	\\										 
\\
\textbf{J2238.4+5902}	&	\textbf{339.61}	&	\textbf{59.05}	&	\textbf{106.55}	&	\textbf{0.47}	&	\textbf{50.41$\pm$11.10}	&	 \textbf{PSR}	&	\textbf{4.53}	&		\textbf{PSR	J2238+5903}	\\
J2239.8+5825	\footnote{This	candidate passes the FDR	cut	$\lambda=0.05$.}	&	339.97	&	58.43	&	106.41	&	-0.16	&	$<\	 $51.39	&	PSR	&	3.01	&		PSR	J2240+5832	\\
\hline
\end{tabular}
\end{table*} 

\begin{table*}
\scriptsize
\centering
\begin{minipage}{130mm}
\caption{Summary of the FDR thresholds and trials corrections for all the candidate lists with different $\lambda$ \label{tab:Summary}}
 \begin{tabular}{lccc}
 \hline
  \hline
Candiate List 	& Pulsar List &  2FGL Extragalactic List  &  TeVCat Extragalactic List \\
\hline
\hline
FDR Threshold with $\lambda$ = 0.1	& 2.08 $\sigma$ 	& 3.45 $\sigma$ 	& 2.23 $\sigma$ \\
Trials Correction with $\lambda$ = 0.1	& 2.73 $\sigma$ 	& 3.63 $\sigma$ 	& 2.72 $\sigma$ \\
\\
FDR Threshold with $\lambda$ = 0.05	& 2.35 $\sigma$ 	& 3.63 $\sigma$ 	& 2.58 $\sigma$ \\
Trials Correction with $\lambda$ = 0.05	& 2.95 $\sigma$ 	& 3.80 $\sigma$ 	& 2.94 $\sigma$ \\
\\
FDR Threshold with $\lambda$ = 0.01	& 3.02 $\sigma$ 	& 4.03 $\sigma$ 	& 3.21 $\sigma$ \\
Trials Correction with $\lambda$ = 0.01	& 3.42 $\sigma$ 	& 4.18 $\sigma$ 	& 3.41 $\sigma$ \\
\\
FDR Threshold with $\lambda$ = 0.001	& 3.66 $\sigma$	& 4.54 $\sigma$ 	& 3.82 $\sigma$ \\
Trials Correction with $\lambda$ = 0.001	& 4.00 $\sigma$ 	& 4.68 $\sigma$ 	& 3.99 $\sigma$ \\
\hline
\end{tabular}
\end{minipage}
\end{table*}

\end{document}